\documentclass[a4paper,fleqn,usenatbib]{mnras}

\usepackage{newtxtext,newtxmath}

\usepackage[T1]{fontenc}
\usepackage{ae,aecompl}

\usepackage{graphicx}	
\usepackage{amsmath}	
\usepackage{amssymb}	

\usepackage{natbib}
\usepackage{booktabs}
\usepackage{longtable, booktabs}
\usepackage[titletoc,title]{appendix}
\usepackage{color}
\usepackage{lscape}

\newcommand{\hb}{H$\beta$}
\newcommand{\ha}{H$\alpha$}
\newcommand{\oiii}{[OIII]}
\newcommand{\oii}{[OII]}

\title[Contribution of clumps and satellites to galaxy mass assembly]{A contribution of star-forming clumps and accreting satellites to the mass assembly of $z \sim 2$ galaxies}

\author[A. Zanella et al.]{
A. Zanella$^{1}$\thanks{E-mail: azanella@eso.org}, E. Le Floc'h$^2$, C. M. Harrison$^1$, E. Daddi$^2$, E. Bernhard$^3$, R. Gobat$^4$,
\newauthor V. Strazzullo$^5$, F. Valentino$^6$, A. Cibinel$^7$, J. S\'anchez Almeida$^{8,9}$, M. Kohandel$^{10}$, 
\newauthor J. Fensch$^1$, M. Behrendt$^{11,12}$,  A. Burkert$^{11,12}$, M. Onodera$^{13,14}$, F. Bournaud$^2$, 
\newauthor J. Scholtz$^{15}$
\\
$^{1}$European Southern Observatory, Karl Schwarzschild Stra\ss e 2, 85748 Garching, Germany\\
$^2$CEA, IRFU, DAp, AIM, Universit\'e Paris-Saclay, Universit\'e Paris Diderot, Sorbonne Paris Cit\'e, CNRS, F-91191 Gif-sur-Yvette, France\\ 
$^3$Department of Physics and Astronomy, University of Sheffield, Sheffield S3 7RH, UK\\
$^4$Instituto de F\'isica, Pontificia Universidad Catolica de Valpara\'iso, Casilla 4059, Valpara\'iso, Chile\\
$^5$Department of Physics, Ludwig-Maximilians-Universitat, Scheinerstr 1, D-81679 Munchen, Germany\\
$^6$Cosmic Dawn Center (DAWN), Niels Bohr Institute, University of Copenhagen, Juliane Maries Vej 30, DK-2100 Copenhagen \O; \\ DTU-Space, Technical University of Denmark, Elektrovej 327, DK-2800 Kgs.\ Lyngby \\
$^7$Astronomy Centre, Department of Physics and Astronomy, University of Sussex, Brighton BN1 9QH, UK\\
$^8$Instituto de Astrof\'\i sica de Canarias La Laguna, Tenerife, Spain\\
$^9$Departamento de Astrof\'\i sica, Universidad de La Laguna, Tenerife, Spain\\
$^{10}$Scuola Normale Superiore, Piazza dei Cavalieri 7, I-56126 Pisa, Italy\\
$^{11}$Max-Planck Institute for Extraterrestrial Physics, Giessenbachstr. 1, D-85748 Garching, Germany\\
$^{12}$Ludwig-Maximilians University Munich, University Observatory, Scheinerstr. 1, D-81679 Munchen, Germany\\
$^{13}$Subaru Telescope, National Astronomical Observatory of Japan, National Institutes of Natural Sciences (NINS), 650 North A'ohoku \\ Place, Hilo, HI 96720, USA\\
$^{14}$Department of Astronomical Science, SOKENDAI (The Graduate University for Advanced Studies), 650 North A'ohoku Place, \\ Hilo, HI, 96720, USA\\
$^{15}$Centre for Extragalactic Astronomy, Department of Physics, Durham University, South Road, Durham DH1 3LE, UK\\
}

\date{Accepted XXX. Received YYY; in original form ZZZ}

\pubyear{2019}

\begin{document}
\label{firstpage}
\pagerange{\pageref{firstpage}--\pageref{lastpage}}
\maketitle

\begin{abstract}
We investigate the contribution of clumps and satellites to the galaxy mass assembly. We analyzed spatially-resolved \textit{Hubble} Space Telescope observations (imaging and slitless spectroscopy) of 53 star-forming galaxies at $z \sim 1$ -- 3. We created continuum and emission line maps and pinpointed residual ``blobs'' detected after subtracting the galaxy disk. Those were separated into compact (unresolved) and extended (resolved) components. Extended components have sizes $\sim$ 2 kpc and comparable stellar mass and age as the galaxy disks, whereas the compact components are 1.5 dex less massive and 0.4 dex younger than the disks. Furthermore the extended blobs are typically found at larger distances from the galaxy barycenter than the compact ones. Prompted by these observations and by the comparison with simulations, we suggest that compact blobs are in-situ formed clumps, whereas the extended ones are accreting satellites. Clumps and satellites enclose respectively $\sim 20$\% and $\lesssim 80$\% of the galaxy stellar mass, $\sim 30$\% and $\sim 20$\% of its star formation rate. Considering the compact blobs, we statistically estimated that massive clumps (M$_\star \gtrsim 10^9$ M$_\odot$) have lifetimes of $\sim$ 650 Myr, and the less massive ones ($10^8 <$ M$_\star < 10^9$ M$_\odot$) of $\sim$ 145 Myr. This supports simulations predicting long-lived clumps (lifetime $\gtrsim 100$ Myr). Finally, $\lesssim$30\% (13\%) of our sample galaxies are undergoing single (multiple) merger(s), they have a projected separation $\lesssim$ 10 kpc, and the typical mass ratio of our satellites is 1:5 (but ranges between 1:10 and 1:1), in agreement with literature results for close pair galaxies.
\end{abstract}




\section{Introduction}
\label{sec:intro}

During the last ten billion years the cosmic star formation rate density has decreased by a factor $\sim 10$ (e.g., \citealt{Lilly1996}; \citealt{Madau1996}; \citealt{Hopkins2006}; \citealt{Madau2014}) and the global stellar mass density has increased by a factor $\sim 2$ (e.g., \citealt{Rudnick2003}; \citealt{Dickinson2003}). 
The mechanisms driving the galaxy mass assembly and evolution through cosmic time are still highly unclear and galaxy-galaxy mergers might play a key role (e.g., \citealt{LeFevre2000}; \citealt{Conselice2009}; \citealt{Cassata2005}; \citealt{Kartaltepe2007}; \citealt{Rawat2008}; \citealt{Bridge2010}; \citealt{Lopez-Sanjuan2009}; \citealt{Lotz2011}).
However, other processes such as cold gas accretion from the cosmic web and subsequent secular evolution are at work (\citealt{Dekel2009}, \citealt{Bouche2010}, \citealt{Dekel2014}, \citealt{Sanchez-Almeida2014}). In particular, in the last decades, spatially-resolved studies of $z \sim 1 - 3$ star-forming galaxies have revealed that they have irregular morphologies dominated by bright knots with blue colors that are generally referred to as ``clumps''. Clumps are star-forming regions and have been studied with multiwavelength datasets, using rest-frame UV and optical continuum data, emission line maps, and CO observations, targeting both field and lensed galaxies (e.g., \citealt{Conselice2004}; \citealt{Elmegreen2005}; \citealt{Elmegreen2007}; \citealt{Genzel2008}; \citealt{Elmegreen2009a}; \citealt{Jones2010}; \citealt{Swinbank2010}; \citealt{Genzel2011}; \citealt{Wisnioski2011};  \citealt{ForsterSchreiber2011}; \citealt{Guo2012b}, \citealt{Livermore2012}; \citealt{Wuyts2012}; \citealt{Murata2014}; \citealt{Tadaki2014}; \citealt{Livermore2015}; \citealt{Genzel2015}; \citealt{Shibuya2016}; \citealt{Mieda2016}; \citealt{Soto2017}; \citealt{Fisher2017a}; \citealt{Dessauges-Zavadsky2017b}; \citealt{Guo2018}; \citealt{Cava2018}). Some works suggest that clumps have typical stellar masses M$_\star \sim 10^7 - 10^9$ M$_\odot$ and sizes $\lesssim 1$ kpc, therefore being 100 -- 1000 times larger and more massive than local star-forming regions (e.g. \citealt{Elmegreen2007}; \citealt{ForsterSchreiber2011}; \citealt{Guo2012b}; \citealt{Elmegreen2013};  \citealt{Soto2017}; \citealt{Dessauges-Zavadsky2018}). However the intrinsic physical properties of clumps continue to be debated. Spatially-resolved studies of high-redshift lensed galaxies have found that clumps have sizes of $\sim 100$ -- 500 pc (e.g. \citealt{Livermore2012}; \citealt{Livermore2015}; \citealt{Cava2018}), up to 10 times smaller than those currently measured in non-lensed galaxies in the same redshift range. It is therefore still unclear whether clumps are single entities or rather clusters of small star-forming regions, blurred into kpc-size clumps due to lack of spatial resolution (\citealt{Behrendt2016}; \citealt{Ceverino2012}). 

Several studies have highlighted that clumps are actively star forming, they typically have high specific star formation rate and star formation efficiency, and resemble small starbursts (e.g. \citealt{Guo2012b}; \citealt{Wuyts2012}; \citealt{Wuyts2013}; \citealt{Bournaud2015}; \citealt{Zanella2015}; \citealt{Mieda2016}; \citealt{Cibinel2017}). Despite their ubiquity at $z \sim 1$ -- 3, contradictory scenarios have been so far proposed to explain the clumps' origin and their evolution. It is not clear whether they are remnants of accreted satellites that have not been completely disrupted by galactic tides (\citealt{Hopkins2013}; \citealt{Puech2009}; \citealt{Puech2010}; \citealt{Wuyts2014}; \citealt{Guo2015}; \citealt{Straughn2015}; \citealt{Ribeiro2017}), or if they are star-forming complexes formed in-situ due to the fragmentation and local collapse of gas-rich, turbulent, high-redshift disks (\citealt{Elmegreen2007}; \citealt{Bournaud2008}; \citealt{Genzel2008}; \citealt{Genzel2011}; \citealt{Guo2012b}, \citealt{Guo2015}; \citealt{Hinojosa-Goni2016}; \citealt{Mieda2016}; \citealt{Fisher2017a}), as predicted by simulations that find high-redshift disks to be gravitationally unstable (\citealt{Noguchi1999}; \citealt{Immeli2004a}, \citealt{Immeli2004b}; \citealt{Bournaud2007}, \citealt{Bournaud2009}; \citealt{Elmegreen2008}; \citealt{Dekel2009}; \citealt{Ceverino2010}, \citealt{Ceverino2012}; \citealt{Dekel2014}; \citealt{Inoue2016}). In particular, if clumps are formed in-situ we should sometimes capture their formation, and hence detect clumps with extremely young age ($\lesssim$ 10 Myr). So far most of the observational studies have been led with broadband imaging, but this alone cannot robustly pinpoint young ages (e.g. \citealt{Wuyts2012}). Spectroscopy, sensitive to gas ionized by very young stars, is needed to probe the earliest clump formation. Only recently some studies have simultaneously used deep continuum and emission line observations to detect young clumps (\citealt{ForsterSchreiber2011}; \citealt{Zanella2015}) and only one clump candidate with age $\lesssim$ 10 Myr (comparable to its free-fall time) has been found so far \citep{Zanella2015}. More observational studies considering simultaneously spatially-resolved imaging and spectroscopy are clearly needed to reach firmer conclusions.

Also the fate of the clumps is debated. Simulation results are contradictory and observations are still uncertain as statistical samples of resolved galaxies are limited and clumps' ages are hard to estimate. Clumps could be quickly disrupted by strong stellar feedback and/or tidal forces that remove the gas and unbound the stellar system. If this is the case, clumps are expected to be short-lived, survive $\lesssim$ 50 Myr (\citealt{Murray2010}; \citealt{Genel2012}; \citealt{Hopkins2012}; \citealt{Buck2017}; \citealt{Oklopcic2017}; \citealt{Tamburello2016}), and they do not affect the structural evolution of the host. Other models instead predict that clumps, given their high star formation efficiencies, transform their molecular gas content into stars in a short timescale and can remain bound, surviving stellar feedback for $\gtrsim 500$ Myr. In this case, due to dynamical friction and gravitational torques, clumps are expected to migrate inward, coalesce, and contribute to the growth of the bulge of the galaxy, possibly feeding the central black hole (\citealt{Bournaud2007}; \citealt{Elmegreen2008}; \citealt{Ceverino2010}; \citealt{Bournaud2011}; \citealt{Gabor2013}; \citealt{Bournaud2014}; \citealt{Mandelker2014}; \citealt{Mandelker2017}). Observational evidences supporting the latter scenario might be the relatively old ages of clumps' stellar populations (age $\gtrsim$ 100 Myr) and the mild negative gradient of clumps' age and/or color with galactocentric distance (i.e. older and redder clumps are preferentially found closer the galaxy barycenter; \citealt{ForsterSchreiber2011}; \citealt{Guo2012b}; \citealt{Shibuya2016}; \citealt{Soto2017}). 
Understanding the clumps' evolution not only could shed light on the mechanisms driving bulge formation, but it could also be key to test the validity of the feedback models used in different simulations (e.g. \citealt{Moody2014}; \citealt{Hopkins2014}; \citealt{Mandelker2017}) and investigate what is the model that reproduces observational results down to the sub-galactic scales of clumps. Finally, investigating clumps formation and physical properties is also a promising way to constrain how galaxies assemble their mass (e.g. mergers and/or secular evolution) and how star formation proceeds at high redshift. 

In this paper we investigate the issues of the origin of the clumps (i.e. disk instability or accretion of satellites) and evolution (i.e. disruption by feedback or survival and inward migration) by using a sample of $z \sim 1 - 3$ star-forming galaxies targeted by deep \textit{Hubble} Space Telescope (\textit{HST}) imaging and slitless spectroscopic observations. 
In Section \ref{sec:observations} we present our data, we discuss the technique that we used to create spatially-resolved emission line maps, we present our final sample of galaxies and their integrated properties. In Sections \ref{sec:clumps_detection} we describe the procedure that we used to find the clumps and satellites, measure their flux and flux uncertainty, and estimate their physical properties and distribution. In Section \ref{sec:results} we present our results, constrain the lifetime of clumps, discuss their inward migration, and report the merger fraction of our sample. Finally, in Section \ref{sec:conclusions} we conclude and summarize. Throughout the paper we adopt a flat $\Lambda$CDM cosmology with $\Omega_m = 0.3$, $\Omega_\Lambda = 0.7$, and $H_0 = 70$ km s$^{-1}$ Mpc$^{-1}$. All magnitudes are AB magnitudes \citep{Oke1974} and we adopt a \cite{Salpeter1955} initial mass function with mass limits 0.1 -- 100 M$_\odot$, unless differently stated.

\section{Data and galaxy sample properties}
\label{sec:observations}

This work is mainly based on \textit{HST} spectroscopic and photometric data taken as part of a project aimed at observing the distant galaxy cluster Cl J1449+0856 at redshift $z = 1.99$ \citep{Gobat2013}. Ancillary data taken with Subaru, Keck, JVLA, APEX, IRAM, ALMA, Chandra, XMM, \textit{Spitzer}, and \textit{Herschel} are also available for most of the sources in the \textit{HST} pointings and were used to characterize the physical properties of our sample. In the following we describe the data used for the analysis, and the techniques adopted to create continuum and emission line maps for our sample galaxies. 

\subsection{\textit{HST} data}
\label{subsec:hst_data}
Spectroscopic and photometric observations targeting the distant galaxy cluster Cl J1449+0856 \citep{Gobat2013} were performed in Cycle 18 (PI: E. Daddi) with \textit{HST} Wide Field Camera 3 (WFC3) using the G141 grism and F140W filter. The imaging was mainly taken to provide information on source positions and morphologies, to correctly model the spectra and facilitate the extraction. The grism observations were executed along three position
angles ($\sim0$, -30, +15 degrees) to correct each spectrum for contamination of nearby sources (Section \ref{subsec:emission_line_maps}), a particularly important task given the high density of sources in the field. The 16 G141 orbits cover a total area of 6.4 arcmin$^2$, with $\sim$ 3 arcmin$^2$ uniformly covered by the three grism orientations. Additional \textit{HST}/WFC3 observations were taken with the F105W and F606W filters (Table \ref{tab:observations}) during Cycle 21 (PI: V. Strazzullo). 

The data were reduced using the aXe pipeline \citep{Kummel2009}. The F140W frames were combined with MultiDrizzle and the resulting image was used to detect the sources and extract the photometry (\citealt{Gobat2011}, \citealt{Gobat2013}, \citealt{Strazzullo2013}). The aXe pipeline processes the grism data and for all the objects in the field of view it creates spectral cutouts calibrated in wavelength and models of the continuum emission, based on input multi-wavelength spectral energy distributions (SEDs, Section \ref{subsec:galaxy_properties}). We processed the spectra taken with different orientations of the grism separately \citep{Gobat2013}. Residual defective pixels not identified by the pipeline (e.g. bad pixels, cosmic ray hits) were removed with the L.A.Cosmic algorithm \citep{vanDokkum2001}.

\subsection{Ancillary data}
\label{subsec:ancillary_data}
A Subaru/MOIRCS near-IR spectroscopic follow-up of 76 sources in the cluster Cl J1449+0856 field has been performed in April 2013. The data have been reduced with the MCSMDP pipeline \citep{Yoshikawa2010} combined with custom IDL scripts \citep{Valentino2015}.

The cluster field has been also followed-up with a large number of
multi-wavelength observations, including photometric data in the U, V
(VLT/FORS), B, R, i, z (Subaru/Suprime-Cam), Y, J, H, K$_\mathrm{s}$ (Subaru/MOIRCS,
plus additional VLT/ISAAC data for J and K$_\mathrm{s}$), F140W, F105W, F606W (\textit{HST}/WFC3), and 3.6, 4.5, 5.8, 8.0$\mu$m (\textit{Spitzer}/IRAC), 24$\mu$m (\textit{Spitzer}/MIPS), 100, 160$\mu$m (\textit{Herschel}/PACS), 250, 350, 500 (\textit{Herschel}/SPIRE) bands, together with Band 3, 4, 7 (ALMA), S, L, Ka (JVLA), and 0.5 -- 10 keV (XMM-Newton), 0.5 -- 8 keV (Chandra).

More details about these ancillary data can be found in \cite{Strazzullo2013}, \cite{Valentino2015}, \cite{Valentino2016}, \cite{Coogan2018}, and references therein.

\begin{table*}
\centering
\caption{\textit{HST}/WFC3 and Subaru/MOIRCS observations extensively used in this study.}
\label{tab:observations}
\begin{tabular}{p{2.5cm} c c c c }
\toprule
\midrule
Instrument             & Date & Time                    & Time      \\
                              &          & (direct imaging)   & (spectroscopy)   \\
                             &          &  (hr)                     &  (hr)        \\
\midrule
\textit{HST}/WFC3              &  2010, 6$^{\mathrm{th}}$ June                & 0.3 (F140W) & 2.7  \\
\textit{HST}/WFC3              &  2010, 25$^{\mathrm{th}}$ June, 1$^{\mathrm{st}}$ July   & 0.6 (F140W) & 7     \\
\textit{HST}/WFC3              &  2010, 9$^{\mathrm{th}}$ July                 & 0.3 (F140W) & 2.7  \\
\textit{HST}/WFC3              &  2013, 20$^{\mathrm{th}}$ May               &  3.3 (F105W)  &  -    \\ 
\textit{HST}/WFC3               &  2013, 20$^{\mathrm{th}}$ May              &  0.3  (F606W)  &  -    \\
Subaru/MOIRCS       & 2013, 7$^{\mathrm{th}}$ - 9$^{\mathrm{th}}$ April                &   -  &  7.3 \\  
\bottomrule
\end{tabular}
\end{table*}

\subsection{Creating spatially-resolved emission line maps}
\label{subsec:emission_line_maps}

We identified sources in the WFC3/F140W band using SExtractor \citep{Bertin1996}. 135 sources were identified in a 6.4 arcsec$^2$ field, 27 of which have been spectroscopically confirmed to be cluster members based on their emission lines (i.e. \oii, H$\alpha$, and/or \oiii) or continuum breaks in the spectral range 1.1 -- 1.7 $\mu$m covered by the WFC3/G141 grism \citep{Gobat2013}. In this work we only focus on the 110 emission line emitters (90 field galaxies and 20 cluster members). For each of these galaxies we considered F140W, F105W, and F606W cutouts probing the stellar continuum of the sources.

Due to the slitless nature of our WFC3 spectroscopic survey, the two-dimensional (2D) light profile of the emission lines is determined by the morphology of the galaxy. Thanks to the high \textit{HST} resolution ($\sim$ 0.1'' -- 0.2'') we can spatially resolve the emission line images and compare their morphology with that of the continuum (namely the F140W, F105W, and F606W cutouts) on kpc-scales.
To this purpose, we created spatially resolved emission line maps, processing each 2D spectrum cutout as follows. First, the overall sky background level was estimated with SExtractor \citep{Bertin1996} and subtracted. Second, we removed the stellar continuum emission of the central object in the cutout (the main target) as well as the contamination introduced by the spectral traces of all potentially surrounding sources, including also higher and lower order dispersion spectra that, given the lack of slits, can overlap with one another. To carry out this step we used the continuum emission models provided by the aXe pipeline for each source in the cutout and we normalized them fitting independently the traces of each object in the cutout. We subtracted the normalized models to the data and we obtained spectral images where only emission lines were left.

Emission lines with full width at half maximum (FWHM) narrower than the spectral resolution (2000 $\mathrm{km\, s^{-1}}$ in our case, thus basically all narrow lines) result in a nearly-monochromatic emission line image of the observed target obtained at the \textit{HST} spatial resolution. For a given galaxy, the detailed morphological structure observed in the imaging (probing the stellar continuum) is not necessarily identical to that visible in spatially-resolved emission line maps. Therefore, it is not possible to construct astrometrically calibrated emission line images directly cross-correlating the spectra and the continuum. For each detected line, emission line maps properly calibrated in astrometry were instead obtained by maximizing the cross-correlation between the spectral images with the three different grism orientations and the continuum probed by the F606W filter. For each grism orientation, in fact, the 2D spectral images of each emission line are identical, the astrometrically aligned spectra differing only for the relative direction of the dispersed continuum. 
Once the relative position of the images that maximizes the cross-correlation has been found, the spectral maps were combined with the IRAF task WDRIZZLE \citep{Fruchter2002}, weighting each single orientation by its corresponding exposure time. The absolute astrometric calibration along the dispersion direction of the grism was determined from the cross-correlation of the [OIII] or H$\alpha$ spectral images (depending on the redshift of the source), as these are the lines detected with the highest signal-to-noise ratio (S/N). The astrometry of the H$\beta$ and [OII] emission maps was afterwards tied to that of the [OIII] or H$\alpha$ maps. For more details about the cross-correlation procedure and the estimate of the associated uncertainties see Appendix \ref{app:cross_correlation}.

Finally, we note that the [OIII]$\lambda$4959,5007\AA~doublet is resolved at the spectral resolution of our data for relatively compact galaxies, but given the fairly small separation in wavelength the
[OIII]$\lambda$4959\AA\ component results in an independent image that is blurred with that of the stronger [OIII]$\lambda$5007\AA\ emission. This produces ghosts with one-third of the [OIII]$\lambda$5007\AA\ flux that are spatially offset in the 2D spectral datasets along directions that are different for each grism orientation. We decided to remove the contribution of [OIII]$\lambda$4959\AA\ in the combined spectral map obtained after the cross-correlation of the three grism orientations, in order to work with higher S/N. We created an effective point-spread function (PSF) of the [OIII] doublet for the combination of our three orientations, which consists of a main lobe corresponding to the 5007\AA\ line and three fainter ones with a flux of $\sim$ 1/9$^{\mathrm{th}}$ of the [OIII]$\lambda$5007\AA\ peak each. With GALFIT \citep{Peng2010} we modelled our combined emission line images using this PSF and finally subtracted the contribution due to the 4959\AA\ lines using the best
fitted model, to obtain cleaned [OIII]$\lambda$5007\AA\ emission line maps. We verified that a similar approach applied instead at each single spectrum orientation would provide entirely consistent results.

\subsection{Final galaxy sample}
\label{subsec:sample}

We started with a parent sample of 135 galaxies identified in the F140W imaging and we only considered the 110 galaxies that were showing at least one emission line ([OII], H$\beta$, [OIII], and/or H$\alpha$) in the 1D spectra, as the goal of this study is to compare the spatially-resolved emission line and stellar continuum maps. 
We excluded from the sample 20 galaxies for which spectra taken with only one or two (but not all three) grism orientations were available, as they were typically at the edge of the WFC3 field of view. We also excluded 20 sources for which the emission line maps were irreparably contaminated either by bad pixels or by the spectral traces of bright nearby sources. Finally, in 16 cases the cross-correlation procedure used to astrometrically calibrate emission line maps (Section \ref{subsec:emission_line_maps}) failed as the 2D emission lines were too faint to reach convergence. Our sample, after the cross-correlation procedure, consists of 54 galaxies. 

We checked for the presence of AGN in our sample galaxies by analyzing our XMM (80 ks, \citealt{Gobat2011}; \citealt{Brusa2005}) and \textit{Chandra} (94 ks, \citealt{Campisi2009}; \citealt{Valentino2016}) data centered on the cluster Cl J1449+0856, covering a total field of view of $\sim 500$ arcmin$^2$.
Only one galaxy (ID607) was detected (L$_{2-10\mathrm{keV}} =  5.2^{+3.4}_{-1.8}\times 10^{43}$ erg s$^{-1}$), suggesting the presence of one X-ray AGN, and we excluded it from our final sample. For the subsample of galaxies that were followed-up with longslit MOIRCS spectroscopy we computed the galaxy-integrated [OIII]/H$\beta$ and [NII]/H$\alpha$ ratios to use the BPT diagnostic diagram \citep{Baldwin1981} to distinguish sources with emission lines powered by star formation from those excited by an AGN. The line ratios measured for our sample galaxies are consistent with being powered by star formation (see the BPT diagram in \citealt{Valentino2015}). For the same subsample of galaxies, we checked for the presence of AGN using the H$\alpha$ equivalent width -- [NII]/H$\alpha$ diagnostics (\citealt{CidFernandes2010}; \citealt{CidFernandes2011}). None of the sample galaxies were selected as AGN according to this diagram \citep{Valentino2015}. Finally, the SEDs of all galaxies are consistent with star formation and do not require additional AGN components \citep{Strazzullo2013}.

Our final sample, after the cross-correlation procedure and the exclusion of one X-ray detected AGN, is therefore made of 53 galaxies and among them 9 are confirmed members of the Cl J1449+0856 cluster. Investigating the effect of the environment on galaxy structure and properties goes beyond the scope of this paper and therefore we do not divide among field and cluster galaxies. We checked however that our results and conclusions would not change if we were to exclude the cluster members from the sample.

\begin{figure}
	\includegraphics[width=0.5\textwidth]{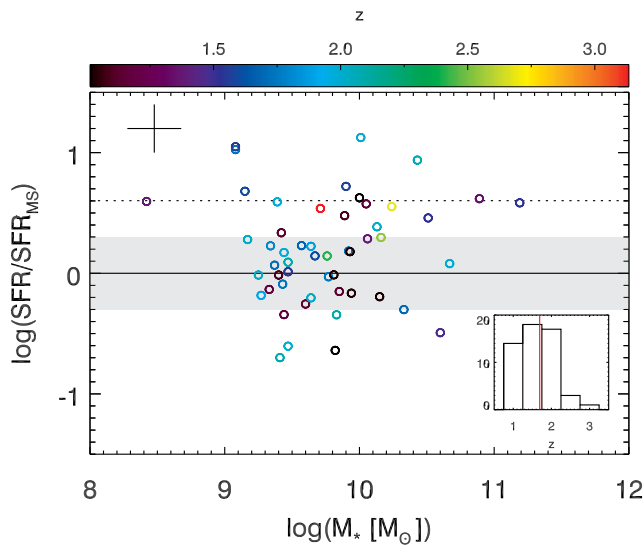}
  \caption{Relation between the stellar mass and the star formation rate, normalized by the star-formation rate of main-sequence galaxies, for our sample. The points are color-coded based on their redshift. Typical stellar mass and star formation rate error bars are indicated in the top-left corner. The star formation rate of main-sequence galaxies is computed considering the redshift and stellar mass of individual sources in our sample, using the relation by \citet{Sargent2014}. According to this relation, z = 2 main-sequence galaxies with M$_\star = 10^{10}$ ($10^9$) M$_\odot$ have SFR $\sim$ 20 (3.5) M$_\odot$ yr$^{-1}$. At z = 1 their SFR, at fixed stellar mass, is $\sim$ 40\% smaller. The dispersion of the mass -- star formation rate relation is reported as the gray area. Starbursts are defined as the sources with SFR $\gtrsim 4\times$ SFR$_{\mathrm{MS}}$ (dashed line). \textit{The inset} in the bottom right corner shows the redshift distribution of our sample and the median redshift is indicated (red line).}
   \label{fig:main_sequence}
\end{figure}

\begin{figure}
	\includegraphics[width=0.5\textwidth]{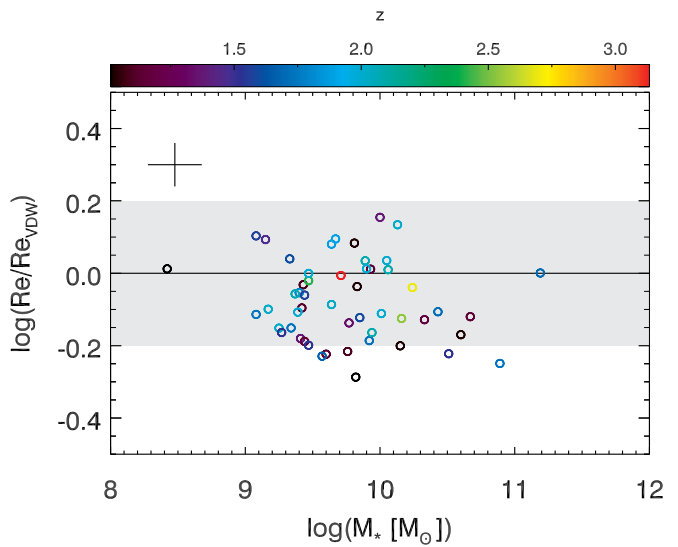}
  \caption{Relation between the mass and size, normalized by the size of typical star-forming galaxies, for our sample. Typical stellar mass and size error bars are indicated in the top-left corner. The points are color-coded based on their redshift. The size of typical star-forming galaxies is computed considering the redshift and stellar mass of individual sources in our sample, using the relation by \citet{vanderWel2014}. According to this relation, z = 2 galaxies with M$_\star = 10^{10}$ ($10^9$) M$_\odot$ have R$_\mathrm{e} \sim$ 2.4 (1.3) kpc. At z = 1 their R$_\mathrm{e}$, at fixed stellar mass, is $\sim$ 30\% larger. The dispersion of the mass -- size relation is reported as the gray area.} 
   \label{fig:mass_size_relation}
\end{figure}

\subsection{Integrated galaxy properties}
\label{subsec:galaxy_properties}

Our final sample is made of 53 galaxies at redshift $z = $ 1.0 -- 3.1 (median redshift $z = 1.7$). 
We determined their properties (i.e. stellar mass, star formation rate, dust extinction) through SED fitting using the FAST code \citep{Kriek2009} on the UV to NIR photometry. \cite{Bruzual2003} stellar population models with constant star formation histories (SFHs), \cite{Salpeter1955} initial mass function (IMF), and \cite{Calzetti2000} extinction law were used. The metallicity was a free parameter of the fit, since in principle at redshift $z \sim 2$ galaxies could have subsolar metallicity, but we checked that fixing the metallicity to the solar value would not change the observed trends. We compared results obtained considering different photometric catalogs: one created with SExtractor based on aperture photometry, and the other one based on GALFIT photometric modeling of the surface brightness of galaxies. When the IRAC photometry suffered from heavy neighbour contamination we excluded the 3.6 $\mu$m -- 4.5 $\mu$m bands from the fitting procedure \citep{Strazzullo2013}. The SED fitting results obtained with the two photometric catalogs were typically consistent (e.g. stellar masses consistent within $\sim$ 0.1 dex). For the subsample of galaxies observed with longslit MOIRCS spectroscopy \citep{Valentino2015} we estimate an average nebular extinction from the Balmer decrement (assuming that H$\alpha/$H$\beta = 2.86$ intrinsically, \citealt{Osterbrok1989}) and we compared it with the extinction derived from SED fitting. The dust attenuation affecting the stellar light ($E(B-V)_{\mathrm{cont}}$, obtained from SED fitting) is typically lower than the one impacting the emission lines ($E(B-V)_{\mathrm{neb}}$, obtained from the Balmer decrement). We therefore used the conversion factor determined by \cite{Kashino2013} to link the two ($E(B-V)_{\mathrm{neb}} = E(B-V)_{\mathrm{cont}}/0.83$). We find that the measurements obtained with the two different methods are consistent within the uncertainties. 

We find that our sample galaxies are typically consistent with the main-sequence of star-forming galaxies estimated by \cite{Sargent2014} at different redshifts, although $\sim$ 10\% of the sample are starbursts (defined as having $\gtrsim 4\times$ enhanced specific star formation rate, Figure \ref{fig:main_sequence}).

We determined the structural parameters (effective radius R$_\mathrm{e}$, S\'ersic index, axial ratio, position angle) of our sample galaxies  by modeling their 2D light profile with GALFIT, using a S\'ersic profile (Section \ref{subsec:finding_blobs}). We find that their sizes, measured from the F140W rest-frame optical band, are consistent with the stellar mass -- size relation of $z \sim 2$ star-forming galaxies reported by \citet[Figure \ref{fig:mass_size_relation}]{vanderWel2014}.
The S\'ersic indices that we find for our sample galaxies are consistent with those of high-redshift disks, having an average S\'ersic index $n \sim$ 1.

\begin{figure*}
\vspace{-4.5cm}
	\includegraphics[width=\textwidth]{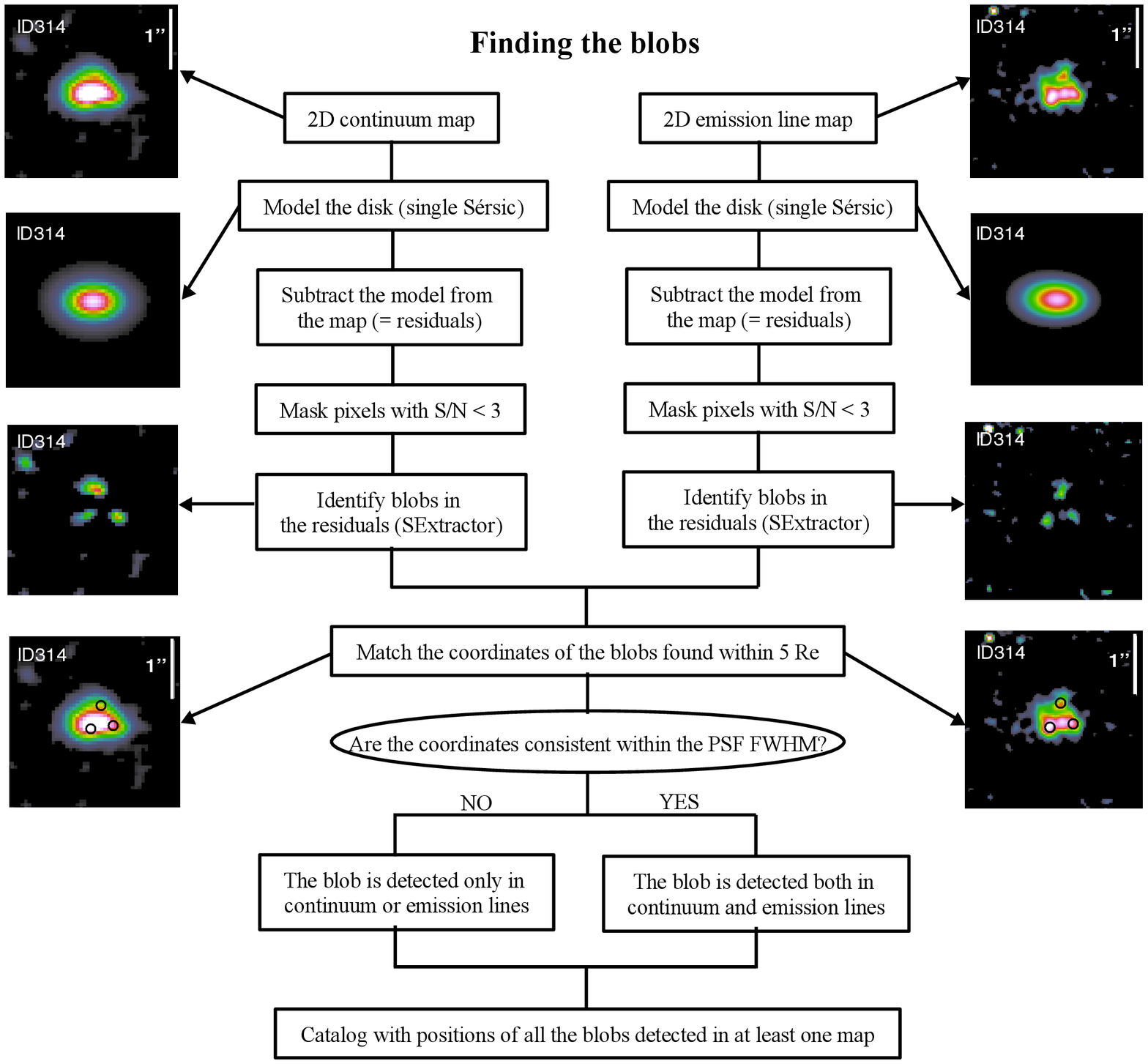}
\vspace{-4.5cm}
  \caption{Flowchart illustrating the procedure that we used to find blobs (Section \ref{subsec:finding_blobs}).}
   \label{fig:flawchart_finding}
\end{figure*}

\section{Continuum and emission line morphological structure}
\label{sec:clumps_detection}

\begin{figure*}
\vspace{-4.5cm}
	\includegraphics[width=\textwidth]{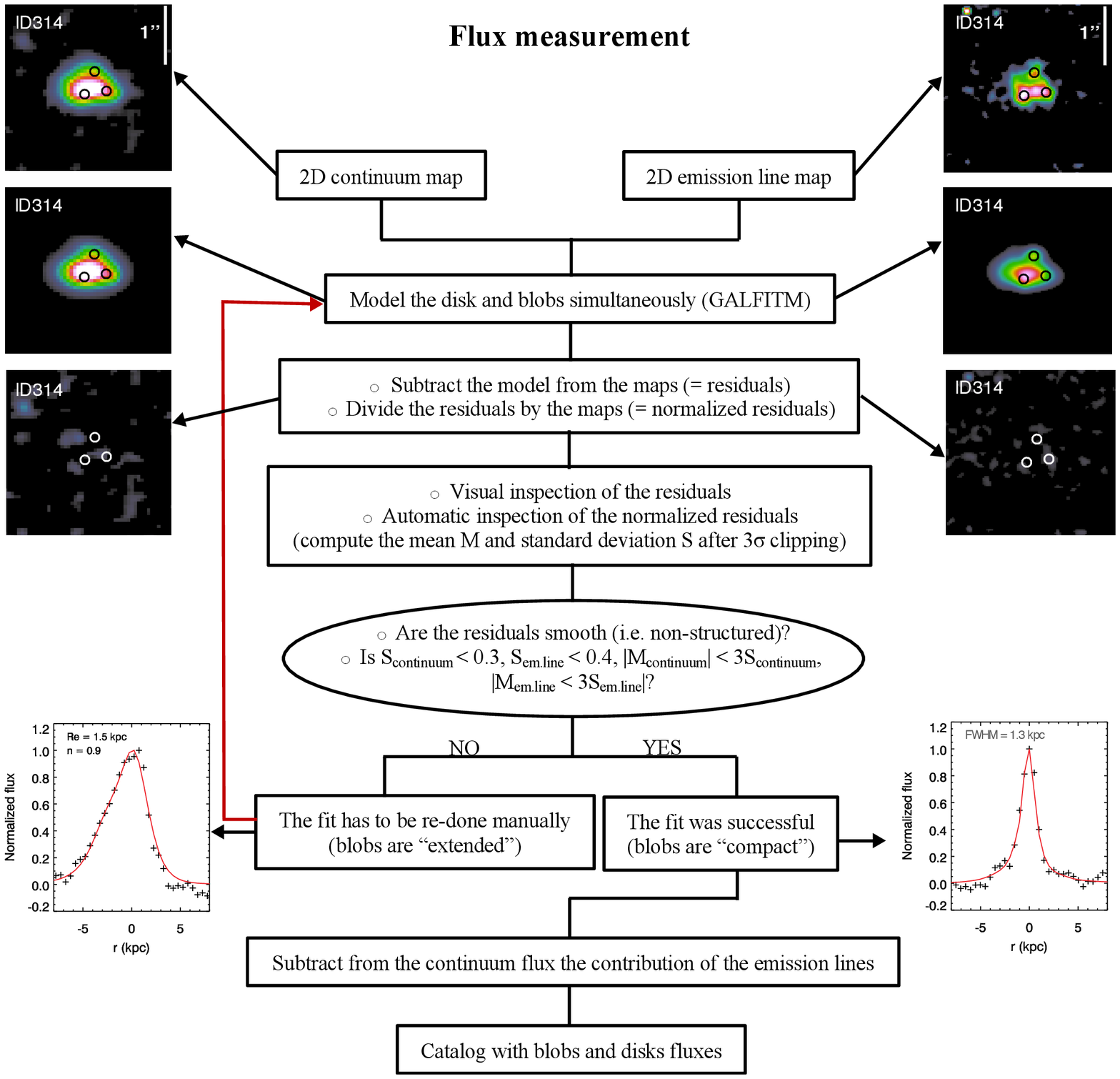}
\vspace{-4.5cm}
  \caption{Flowchart illustrating the procedure that we used to measure the flux of the blobs and the underlying disks. S$_\mathrm{continuum}$ (S$_\mathrm{em.line}$) and M$_\mathrm{continuum}$ (M$_\mathrm{em.line}$) are resepctively the standard deviation and mean of the distribution of the normalized residuals after 3$\sigma$ clipping for the continuum (emission line) maps (Section \ref{subsec:blobs_flux}).}
   \label{fig:flawchart_measuring}
\end{figure*}

In this work we investigate the morphological structure of galaxies as probed by their spatially-resolved, continuum and emission line maps. Most of our galaxies show a diffuse, disk-like stellar continuum emission plus some irregular structures (e.g. star-forming regions, clumps, merging satellites). In this Section we call ``blobs'' all the significantly detected residuals that depart from the diffuse stellar disk, similarly to \cite{Guo2018}. In the following we describe the method that we used to deblend the blobs from the underlying disk (Figure \ref{fig:flawchart_finding}), how we estimated their continuum and emission line fluxes (Figure \ref{fig:flawchart_measuring}), determined their observed and derived physical properties, and fitted their distributions.

\subsection{Finding the blobs}
\label{subsec:finding_blobs}

To detect blobs in our sample galaxies and disentangle their emission from that of the underlying diffuse disk, we created the following automated procedure (Figure \ref{fig:flawchart_finding}). First, we modelled with GALFIT \citep{Peng2010} the 2D light profile of the continuum and emission line maps independently. We adopted a single S\'ersic profile and we subtracted the best-fit model from each map. In this way, we verified whether the galaxy could be considered as a smooth disk, or if additional blobs were showing up in the residuals, after the disk subtraction. To identify additional blobs, we ran SExtractor separately on the broad-band and emission line residuals, after masking the pixels with a S/N lower than 3, to limit the number of spurious detections. We matched the coordinates of the blobs that we found in the continuum and emission line maps, within 5 R$_\mathrm{e}$ from the galaxy barycenter to include in the sample both star-forming regions belonging to the disk and close accreting satellites. We considered that two detections were matched if their offset in the broad-band images with respect to the emission line maps was less than the FWHM of the point-spread function (PSF) of the F140W image ($\lesssim 0.15$''), the band with the lowest resolution. Matching the coordinates of the blobs was a necessary step as small misalignments between the broad-band images and the spectral maps might still have been present even after the cross-correlation procedure that we applied to calibrate the astrometry (Section \ref{subsec:emission_line_maps}). We found that the average offset between continuum and emission line maps is smaller than 0.03'', consistent with the distorsions that we estimated with the cross-correlation procedure (Appendix \ref{app:cross_correlation}).
We considered also blobs that were detected in the continuum but not in the emission line maps, and viceversa, not to bias our study and we created a catalogue with the coordinates of all the 98 blobs that have been identified by SExtractor (Table \ref{tab:properties_blobs}).
We checked that running this process iteratively (i.e. fit the disk, find the blobs in the residuals, mask them, fit the disk again and look for residual blobs) does not improve our completeness or detection limits, so we only adopted one iteration.

\subsection{Continuum and emission line flux measurements}
\label{subsec:blobs_flux}

After finding the blobs in our sample, we estimated the flux of their continuum and line emission as follows and illustrated in Figure \ref{fig:flawchart_measuring}. We fitted again the 2D light profiles of our sample galaxies, this time simultaneously considering a S\'ersic profile to model the diffuse disk component plus PSFs at the location of the blobs found with SExtractor (Section \ref{subsec:finding_blobs}). To this aim we used the fitting algorithm GALFITM \citep{Vika2013}, considering as initial guesses for the location of the model PSFs the coordinates of the blobs detected by SExtractor (Section \ref{subsec:finding_blobs}). We used PSF profiles to model the blobs since star-forming regions at $z \sim 1$ -- 2 are expected to be unresolved at the \textit{HST} resolution ($\sim 1$ kpc, e.g. \citealt{Guo2015}). GALFITM is an algorithm that allows the user to simultaneously fit multiple images of the same galaxy taken at different wavelengths. We used it to model simultaneously the F140W, F105W, and F606W direct images, together with the available emission line maps of each galaxy. The main advantage of this algorithm with respect to GALFIT is the fact that it is possible to force all the components of the model (e.g. diffuse S\'ersic profile and PSFs, in our case) to keep the same relative distances, while the whole model can rigidly shift from one band to another to fit the observations, even if residual minor misalignments between continuum and emission line maps are present. We visually inspected the residuals of every galaxy, after the subtraction of the best fit model, to check for the reliability of the fits. For some galaxies the procedure did not succeed, leaving non-negligible and/or structured residuals. In these cases ($\sim$ 30\% of the sample) we had to include an off-nuclear S\'ersic profile instead of a PSF at the location suggested by SExtractor. We further checked our results in two ways. First, we verified that the disks' center coordinates as determined by GALFITM in the F140W images were consistent with those of the barycenter estimated with SExtractor on the same maps. Second, we compared the location of the disks' center coordinates estimated by GALFITM in the emission line maps with those obtained in the continuum maps. We found that the models are offset less than 0.03'', without any systematic trend, completely consistent with the effects expected due to distorsions (estimated to be at maximum 0.06''). Finally, we verified the reliability of the disks effective radius determined by GALFITM by looking at the mass -- size relation of our sample galaxies (Figure \ref{fig:mass_size_relation}, Section \ref{subsec:galaxy_properties}). 
We show the results of the fits of our sample galaxies' light profiles in Appendix \ref{app:fig_stat}.

\subsubsection{Continuum correction for emission line contributions}
\label{subsec:emission_line_contributions}

At the redshift of our sources, the F140W bandpass includes the \oiii ~ doublet and H$\beta$, and the F105W bandpass includes the \oii ~ doublet. It is therefore necessary to account for the nebular line emission when studying the morphology of the broad-band data and when determining the stellar continuum flux of the blobs. After estimating the emission line flux of each blob, we computed the contribution of the measured \oiii ~ and H$\beta$ to the F140W and of \oii ~ to the F105W, considering the transmission function of each filter. We then subtracted the contribution of the nebular emission from the continuum flux estimated for each blob. In general, the nebular emission contributes $\lesssim$ 25\% of the integrated F140W flux and $\lesssim$ 10\% of the integrated F105W flux. There are a few extreme cases though, where the \oiii ~ and H$\beta$ flux make up $\sim$ 100\% of the F140W flux (e.g. some blobs hosted by ID568, ID843, and ID834). In these cases the blobs have a stellar continuum flux that is likely lower than the limiting magnitude of our observations and the detection in the broad-band images is almost entirely due to the nebular emission. For these blobs we estimated a 3$\sigma$ upper limit of the continuum flux as detailed in Section \ref{subsec:blobs_uncertainties}. In the following we always refer to nebular line emission-corrected continuum fluxes. We did not correct the fluxes for the nebular continuum emission as this is a negligible contribution ($<$ 20\% at the wavelengths considered in this study, for the metallicity range spanned by our blobs, for stellar populations with ages $\gtrsim 5$ Myr, e.g. \citealt{Byler2017}) and it would increase the uncertainties without changing our conclusions.

\subsection{Estimate of the flux uncertainties and sample completeness}
\label{subsec:blobs_uncertainties}

To estimate the uncertainties associated to our flux measurements we used 1000 Monte Carlo simulations. The main issue we wanted to understand with these simulations was how well GALFITM retrieved PSFs on top of a disk. We injected one fake PSF at the time, with magnitude randomly chosen in the range 24 -- 31 mag (motivated by the range of magnitudes of our observed blobs), with random locations on top of the observed continuum and emission line maps. We took care not to inject fake PSFs on top of already existing blobs, imposing a minimum distance from any detected blob of about 0.15'' (equal to the FWHM of the PSF in the F140W image), although this might be a realistic test of blending of nearby blobs. We treated these simulated images with the same procedure reported in Section \ref{subsec:blobs_flux}, running GALFITM simultaneously on the various continuum and emission line maps. To determine the uncertainties associated to the flux of the blobs, we divided the simulated PSFs in bins based on their distance from the galaxy barycenter and the luminosity contrast between the PSF and the underlying disk (L$_\mathrm{PSF}$/L$_\mathrm{disk}$) as measured by GALFITM on the real data. This step was needed since the ability of GALFITM to correctly retrieve the flux is highly dependent on the contrast of the PSF with respect to the underlying diffuse disk. For each bin, we computed the difference between the known input flux of the fake injected PSF and the one retrieved by GALFITM. The standard deviation of the sigma clipped distribution of these differences gave us, in each contrast and distance bin, the flux uncertainty. A detailed description of the procedure we used can be found in Appendix \ref{app:uncertainties}.

For each observed blob, given the flux estimated with GALFITM and its associated uncertainty estimated with the method described above, we computed the S/N. We considered as detections only the blobs with S/N $\geq 3$. If in a given band the S/N was $< 3$, we adopted a 3$\sigma$ upper limit based on the estimated uncertainty. Only the blobs that were detected with S/N $\geq$ 3 in at least one map were retained. We note that it was important to keep in our catalogue also the blobs that had detected emission lines, but not continuum, in order to study very young star-forming regions (e.g. our analysis of Vyc1, \citealt{Zanella2015}). We also retained in our sample blobs that were detected in the continuum, but only had an upper limit for the emission lines, since they could be very old star-forming regions or satellites (see the figures in Appendix \ref{app:fig_stat} for examples).

To estimate the flux completeness of our sample of blobs, we used the same Monte Carlo simulations described above. 
We concluded that our sample is 50\% complete down to 28 mag (25.9 mag arcsec$^{-2}$) in F140W, 28.3 mag (26 mag arcsec$^{-2}$) in F105W, 28.1 mag (25.9 mag arcsec$^{-2}$) in F606W imaging, and 28.8 mag (26.8 mag arcsec$^{-2}$) in emission line maps. The completeness of our sample decreases when the contrast between the luminosity of the blob and that of the underlying disk decreases, and when the blobs are closer to the galaxy barycenter (Figure \ref{fig:completeness_functions}).

\subsection{Observed properties of the blobs}
\label{sec:blobs_obs_prop}

By simultaneously analyzing spatially-resolved continuum and emission line maps of a sample of 53 galaxies at $z \sim 1$ -- 3 we identified residual components after subtracting the galaxy disk. In our sample, 30\% of the galaxies have a single S\'ersic light profile and the remaining 70\% show additional substructure (``blobs''). In this Section we discuss how we estimated the observed properties of the blobs such as their light profile and size, galactocentric distance, continuum and emission line luminosity, and equivalent width. We report the properties of the blobs in Table \ref{tab:properties_blobs}. 

\subsubsection{Compact and extended components}
\label{subsec:light_profile}

\begin{figure}
	\includegraphics[width=0.45\textwidth]{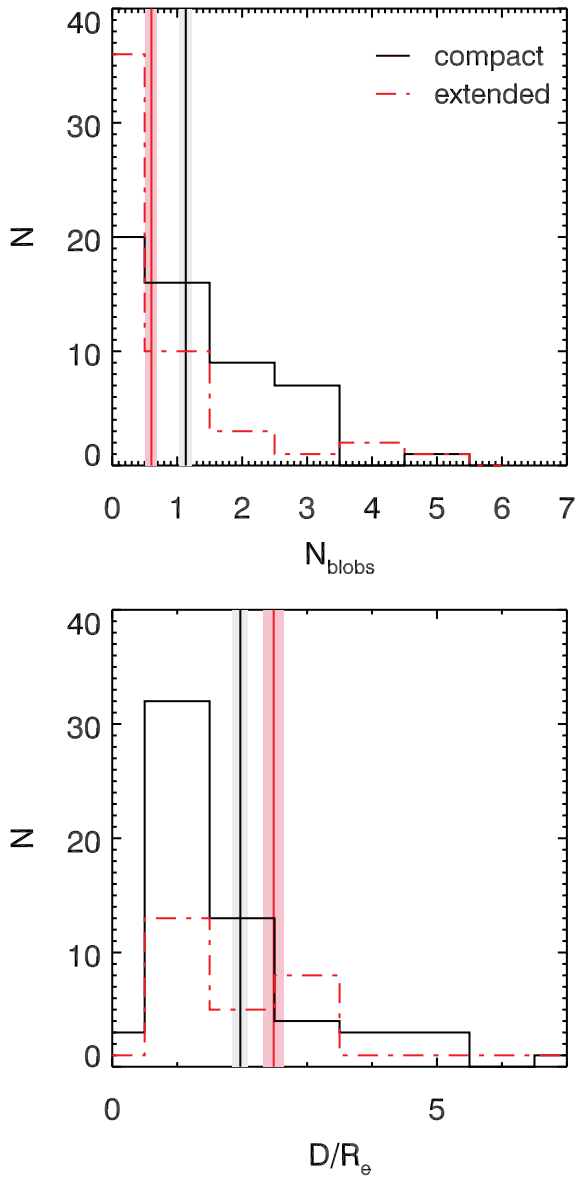}
  \caption{Distribution of the number of blobs per galaxy and their galactocentric distance. Compact blobs (with PSF profile) are shown in black, extended ones (with S\'ersic profile) in red. \textit{Top panel:} number of blobs per galaxy. We also report in the histogram the galaxies that do not have compact (extended) blobs either because they only show a disk component or because they only host extended (compact) blobs. \textit{Bottom panel:} deprojected distance of each blob from the galaxy barycenter normalized by the effective radius of the galaxy stellar disk. We also report the mean (vertical red and black lines) and 3$\sigma$ uncertainties of the mean (thickness of the gray and light red bands) of each distribution.}
   \label{fig:distr_extra}
\end{figure}

When fitting the 2D light profile of galaxies we used a S\'ersic component to reproduce the galaxy disk, and PSFs to fit the blobs identified in the residuals (Section \ref{subsec:blobs_flux}). This procedure was successful for the majority of the galaxies in our sample. However, in $\sim$ 30\% of the cases the best-fit model was not satisfactory as we would obtain structured and non-negligible residuals. The main reason for this was the poor fit of the blobs and the issue was solved by using a S\'ersic profile instead of a PSF to model their light profile. In our final sample, 66 blobs ($\sim$ 70\% of the sample) have a PSF-like profile (and are therefore unresolved at the resolution of \textit{HST}) and 32 have a S\'ersic profile (Figure \ref{fig:distr_extra}).

The blobs with S\'ersic profile have a median effective radius R$_\mathrm{e} =$ 2 kpc (whereas the PSF of our observations has a FWHM $\sim$ 1.3 kpc) and a median S\'ersic index $n = 1.1$. 

In the following we keep separate these two populations of blobs and compare their physical properties to gain insights on their nature. We refer to blobs with PSF-like profile as ``compact'' and to those with S\'ersic profile as ``extended''.

In our sample, 70\% of the galaxies host at least one blob and the average number of blobs per galaxy is 1.8 $\pm$ 0.1 (1.1 $\pm$ 0.1 are compact and 0.6 $\pm$ 0.1 are extended). The standard deviation of the distribution is 0.2 (for both extended and compact blobs). Extended blobs are on average found at larger deprojected distance (d$_\mathrm{depr} \sim$ 2.1 $\pm 0.05$ R$_\mathrm{e}$) from the galaxy barycenter than compact ones (d$_\mathrm{depr} \sim$ 1.3 $\pm 0.04$ R$_\mathrm{e}$). Both distributions have a standard deviation of $\sim$ 0.3 R$_\mathrm{e}$. We report in Figure \ref{fig:distr_extra} the distribution of number of compact and extended blobs per galaxy (top panel) and of their distance from the galaxy barycenter (bottom panel). We also performed a two-sided Kolmogorov-Smirnov test to estimate the probability that the distribution of the number of compact and extended blobs per galaxy are drawn from the same parent distribution. We obtained a P-value of 0.02 and can therefore reject the null hypothesis. Similarly, for the distribution of distances from the galaxy barycenter, the Kolmogorov-Smirnov test gives a P-value of $\sim 10^{-6}$, indicating that the properties of compact and extended blobs are significantly different and likely they are not drawn from the same parent distribution.

\subsubsection{Galactocentric distance}
\label{subsec:galactocentric_distance}

We defined the observed (projected) distance of the blobs from the nucleus of the host galaxy as the difference between the coordinates of each blob and the center of the diffuse component determined by GALFITM. Considering the axial ratio ($q$) of the disk measured by GALFITM and the angular distance of each blob from the galaxy major axis ($\phi$), we computed the deprojected distance of the clumps from the nucleus as
\begin{equation}
d_\mathrm{depr} = \sqrt{(d \cos \phi)^2 + ((d \sin \phi)/q)^2}
\end{equation}

\noindent
where $d$ is the observed (projected) distance of the blob from the galaxy barycenter.

The deprojected distance of blobs from the galaxy nucleus assumes that the galaxy is axi-symmetric and it depends on the inclination and position angle of the galaxy itself. Whereas these assumptions are at first order correct if the blobs are embedded in the galaxy disk, this might not be appropriate if the blobs are satellites that do not belong to the disk. The projected distance of potential satellites in our sample (Section \ref{subsec:clumps_satellites}) differs on average $\sim 20$\% from the deprojected one and considering it instead would not change our main conclusions.

\subsubsection{Continuum and emission line luminosity}
\label{subsec:luminosities}

We estimated the intrinsic continuum and emission line luminosity of the blobs by correcting the observed flux (Section \ref{subsec:blobs_flux}) for the effect of dust extinction. We considered that the reddening affecting the blobs is the same as the average one measured for the whole galaxy (Section \ref{subsec:galaxy_properties}), as we could not perform spatially resolved SED fitting with the available data. This assumption, at least for the star-forming regions that are part of our sample ($\sim 70$\% of our blobs sample), is supported by literature works showing that clumps and their parent galaxies typically are affected by comparable extinction (\citealt{Elmegreen2007}; \citealt{Wuyts2013}), and by our previous results obtained for Vyc1 \citep{Zanella2015}. Part of our blobs sample ($\sim 30$\%) instead is likely made of merging satellites (Section \ref{subsec:clumps_satellites}) for which the assumption that they are affected by the same dust extinction as the host galaxy  is more uncertain. However most of these satellites seem to have comparable stellar mass as the host galaxy (Section \ref{subsec:properties_blobs}) and therefore we do not expect this assumption to systematically affect the properties (e.g. luminosity, SFR, age) of the satellites, but possibly to increase their scatter, hence our conclusions should not be systematically biased.

We estimated the uncertainties on the blobs intrinsic luminosity by considering the uncertainties associated to their observed flux (Section \ref{subsec:blobs_uncertainties}) and those associated to the reddening estimate. Upper limits on the observed continuum or emission line flux are translated into upper limits on the intrinsic luminosity.

\subsubsection{Equivalent width}
\label{subsec:equivalent_width}

We estimated the equivalent width (EW) of each blob in the sample, as the ratio of the emission line flux and the flux of the continuum close in wavelength to the considered emission line. To determine the \oiii , H$\beta$, and H$\alpha$ equivalent widths we used the continuum estimated from the F140W image, whereas for the \oii ~ equivalent width we used the continuum estimated from the F105W map. In the following we refer to rest-frame equivalent widths. 

We estimated the uncertainties on the equivalent width by propagating the uncertainties on the continuum and emission line flux, and those on the reddening. An upper limit on the continuum (emission line) flux is translated into lower (upper) limit on the equivalent width.

\vspace{0.5cm}
Similarly, we also estimated the continuum and emission line luminosity, and the equivalent width of the underlying galaxy disks, with associated uncertainties and/or upper/lower limits.

\subsection{Derived physical properties of the blobs}
\label{sec:blobs_physical_prop}

Given the observables described in the previous Section, we estimated the physical properties of the blobs (stellar mass, star formation rate, specific star formation rate, age, metallicity) and their uncertainties. We discuss in the following the method we used to derive each property.
We report the properties of the blobs in Table \ref{tab:properties_blobs}. 

\subsubsection{Stellar mass}
\label{subsec:stellar_mass}

We estimated the stellar mass of the blobs by multiplying the mass-to-light ratio (M/L) of the host galaxy by the luminosity of the continuum emission of the blobs, as measured from the F140W images. This scaling assumes that the M/L ratio remains constant across the galaxy disk and it does not take into account the fact that the blobs could have different colors with respect to the host galaxy, mainly due to their younger age. Furthermore differences in the reddening and star formation history of the blobs with respect to the underlying galaxy disk are expected to affect the mass-to-light ratio. To correct for these possible effects, we considered the relation between J-band and H-band colors and the M/L ratio found by \cite{Forster-Schreiber2011b}. By comparing the color of our sample blobs (determined using the F105W and F140W continuum maps, Appendix \ref{app:mass_maps}) with those of the underlying disks, we estimated the M/L ratio correction needed to properly estimate the stellar mass of the blobs. In case the blobs are not detected in the continuum ($\sim 20$\% of our sample) and only upper limits on their color are available, we estimate their mass by simply considering the M/L ratio of the host disk (without further corrections).
If we were not to correct our mass estimates for the different colors of blobs and disks, we would obtain on average $\sim 0.15$ dex larger (smaller) masses for the compact (extended) blobs and our conclusions would still hold.

We estimated the uncertainties on the stellar mass of the blobs by considering the uncertainty on their continuum luminosity and a typical uncertainty of $\sim$ 0.3 dex on the M/L ratio. Upper limits on the continuum flux are translated into stellar mass upper limits.

\subsubsection{Star formation rate}
\label{subsec:sfr}

We estimated the star formation rate of our sample blobs in multiple ways, depending on the available emission lines. In case the H$\alpha$ line was detected we used the following equation: SFR $= 7.9\times 10^{-42}$ L$_{\mathrm{H\alpha}}$ \citep{Kennicutt1998}, where L$_\mathrm{H\alpha}$ is the intrinsic H$\alpha$ luminosity. In case the \oii ~  was detected we still used the previous equation, considering an intrinsic ratio \oii /H$\alpha = 1$ \citep{Kewley2004}. In case the \oiii ~ was detected we used the following relation: SFR $= ($L$_\mathrm{\oiii}\times 10^{-41.39})^{1.47}$ \citep{Valentino2017}, where L$_\mathrm{\oiii}$ is the intrinsic \oiii ~luminosity. We did not estimate the SFR using the H$\beta$ line since its S/N was generally too low to obtain reliable results. When possible we compared the SFR obtained, for the same galaxy, using multiple lines. These SFR estimates are typically within $<$ 0.2 dex. Furthermore, we compared the SFR estimated from emission lines with that obtained considering the F606W continuum flux (probing the rest-frame UV continuum at the redshift of our sources) and the relation SFR $= 1.4\times 10^{-28}$L$_\mathrm{\nu}$, where L$_\mathrm{\nu}$ is the rest-frame UV intrinsic luminosity \citep{Kennicutt1998}. They are in good agreement (better than 0.2 dex). However, we prefer to use the SFRs estimated using the emission lines, as they typically have higher S/N than the F606W continuum flux. In the following, for each blob, we consider the SFR estimated using the emission line with the highest S/N.

We estimated the SFR uncertainties by considering the uncertainties associated to the emission line luminosity and to the coefficients of the relation used to convert luminosity into SFR. Upper limits on the emission line flux are translated into upper limits on the SFR.

\subsubsection{Specific star formation rate and distance from main-sequence}
\label{subsec:ssfr}

We estimated the specific star formation rate of blobs and diffuse disks from their star formation rate and stellar mass: sSFR = SFR/M$_\star$. Uncertainties on the sSFR were estimated by propagating the errors on SFR and M$_\star$. Upper limits on the SFR (M$_\star$) give upper (lower) limits on the sSFR.

By considering the main-sequence of star-forming galaxies determined by \cite{Sargent2014} at different redshifts, we also estimated the ratio of sSFR of blobs (and disks) and that of a main-sequence source with the same redshift and stellar mass $\Delta_{\mathrm{MS}} = \log($sSFR/sSFR$_{\mathrm{MS}})$. This indicates whether blobs (and disks) lie on the main-sequence or have enhanced/decreased sSFR. Uncertainties on $\Delta_{\mathrm{MS}}$ were estimated based on the uncertainties on the sSFR and the scatter of the main-sequence determined by \cite{Sargent2014}. Upper (lower) limits on the sSFR are translated into upper (lower) limits on $\Delta_{\mathrm{MS}}$.

\subsubsection{Age}
\label{subsec:blobs_age}

The equivalent width (EW) is almost insensitive to dust extinction, if the emission line and continuum originate from the same region, but it strongly varies with the stellar age of a stellar population. It is therefore a good tool to constrain the stellar ages of the blobs. 
Hence to constrain the age of the blobs we used the tight correlation between their equivalent width and the evolution of their stellar population. We considered Starburst99 stellar population synthesis models to compute the evolution of the H$\alpha$ and H$\beta$ EW as a function of the age of the stellar population. We considered a \cite{Salpeter1955} IMF, the average metallicity of the galaxy estimated through line ratios (Section \ref{subsec:metallicity}), when available, or SED fitting (Section \ref{subsec:galaxy_properties}), and two different star formation histories (SFHs): a constant star formation law and a SFH obtained from our hydrodynamical simulations (\citealt{Bournaud2014}, \citealt{Zanella2015}), characterized by a burst of star formation lasting for almost 20 Myr, followed by a rapid decline. To estimate the evolution of the [OIII] and [OII] emission lines that are not directly modelled by Starburst99, we rescaled respectively the H$\beta$ models considering the typical [OIII]/H$\beta$ line ratio for star-forming galaxies at $z \sim 2$ \citep{Steidel2014}, and the H$\alpha$ models assuming an intrinsic ratio H$\alpha$/[OII] $= 1$ \citep{Kewley2004}. The blobs seem to have a comparable metallicity as the host disks (Figure \ref{fig:trends}) and therefore assuming that blobs and their host galaxies have similar line ratios is reasonable. Future deep spatially-resolved observations of sample of blobs (e.g. with VLT/ERIS, \textit{JWST}/NIRSpec) will be key to further compare the line ratios of blobs and galaxies and confirm this assumption. Comparing these model predictions with the measured EW, we estimated the age of the blobs. We compared the ages determined assuming the two SFHs mentioned above and we checked that they are consistent within the error bars. In the following analysis we will consider the age estimated assuming a SFH with constant SFR. Changing the metallicity by 1.5 dex, varying the reddening by 0.2 dex, or changing the IMF from Salpeter-like to top-heavy would change the predicted equivalent width of a 10 -- 1000 Myr old stellar population by $\lesssim$ 0.1 dex.
By using a similar procedure we also estimated the age of the underlying disks. Despite the fact that this method is more accurate for young stellar populations (age $\lesssim$ 100 Myr), we could put lower limits to the age of the disks that are consistent with the estimates obtained from the SED fitting of the integrated galaxy photometry. We notice however that the equivalent width is also sensitive to the specific star formation rate since EW = f$_{\mathrm{line}}/$f$_{\mathrm{continuum}} \sim$ SFR/M$_\star =$ sSFR. Therefore the age and sSFR estimates of the blobs are not independent.

We estimated the uncertainty on the age of the blobs considering the uncertainties on the equivalent width and a 0.1 dex uncertainty on the models. Upper limits on the emission line (continuum) flux give lower (upper) limits on the blob's age.

\subsubsection{Metallicity}
\label{subsec:metallicity}

For galaxies in the redshift range $z \sim$ 1.9 -- 2.4, both the \oiii ~and \oii ~emission lines were in the spectral range covered by the G141 grism. We estimated the gas-phase metallicity of these blobs from the reddening-corrected \oiii /\oii ~emission line ratio, using the calibration by \cite{Maiolino2008}. In these cases we also estimated the metallicity of the disks considering the same emission line ratio and calibration.
The \oiii/\oii ~ratio seems also to be sensitive to the ionization parameter and the metallicity calibration by \cite{Maiolino2008} implicitly assumes that the ionization parameter of high-redshift galaxies (and blobs, in our case) is similar to that of $z = 0$ sources. It is still unclear whether this assumption is correct (\citealt{Onodera2016, Sanders2016}), but recently direct metallicity measurements through the \oiii 4363 emission line have shown that the metallicity calibration from \cite{Maiolino2008} seems to hold even for $z \gtrsim 1$ star-forming galaxies (\citealt{Jones2015}, \citealt{Sanders2016}). It remains to be tested whether this assumption also holds for high-redshift blobs, where the ionization conditions could be different from those in local star-forming regions.

The uncertainties associated to the metallicity estimates are derived considering the uncertainties on the \oiii ~and \oii ~flux, the reddening, and the calibration by \cite{Maiolino2008}.

\vspace{0.5cm}
The procedure that we used to estimate the physical properties of the underlying galaxy disks, with associated uncertainties and/or upper/lower limits, is the same that we adopted for the blobs.

\begin{figure*}
	\includegraphics[width=\textwidth]{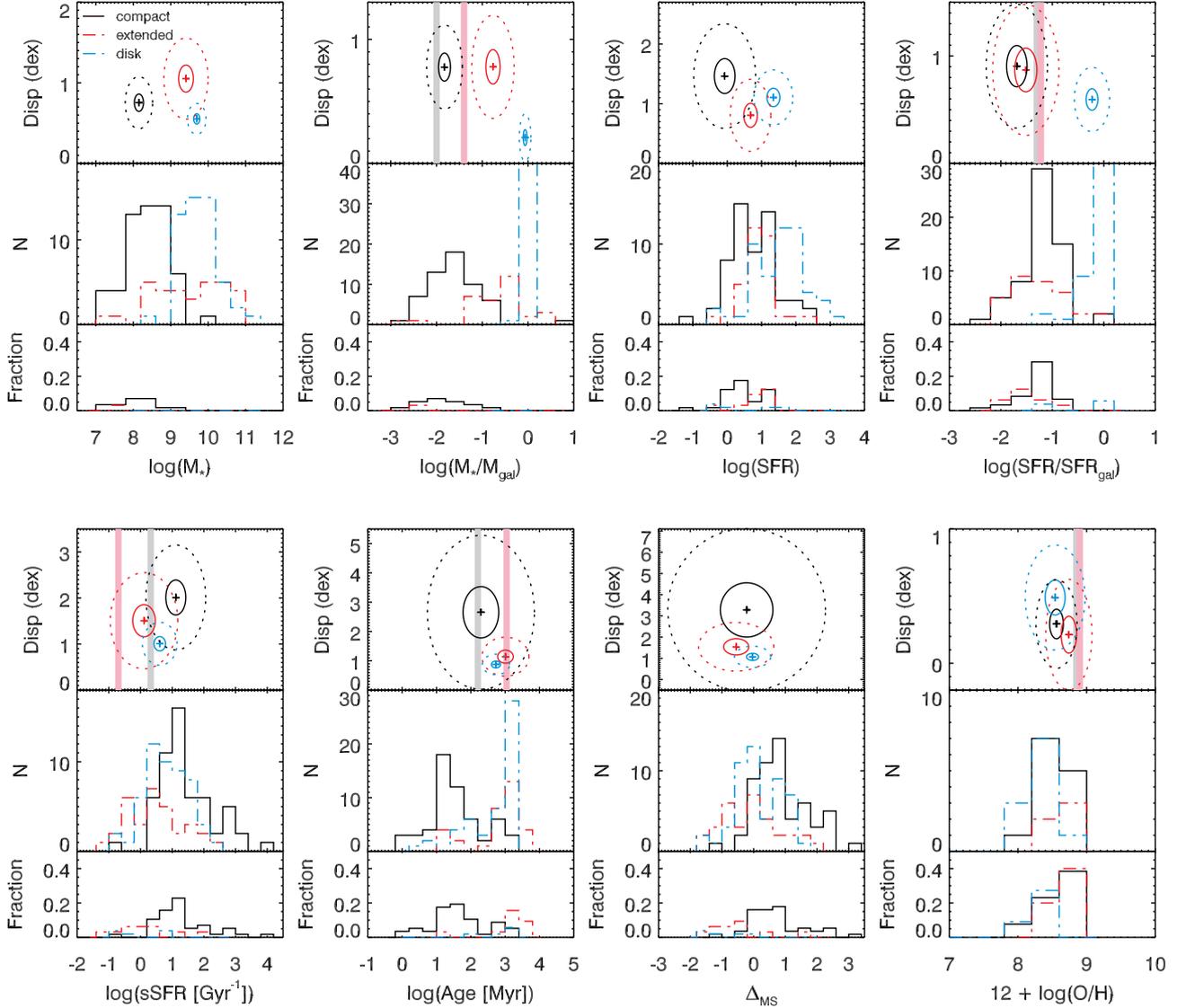}
  \caption{Distributions of the physical properties of blobs and disks. Compact blobs are shown in black, extended ones in red, and the underlying disks in cyan. \textit{Top panels (from left to right):} we show the distribution of stellar mass, stellar mass normalized by the total (blobs plus underlying disk) galaxy mass, star formation rate, and star formation rate normalized by the total galaxy star formation rate. \textit{Bottom panels (from left to right):} we show the distribution of specific star formation rate, stellar age, distance from the main-sequence of star-forming galaxies, and gas-phase metallicity. For each physical property (e.g. stellar mass) we show in the \textit{middle panel} the distribution of the parameters, in the \textit{bottom panel} the fraction of upper or lower limits in each bin, and in the \textit{top panel} the mean and dispersion (cross) of each distribution (solid and dotted ellipses indicate the 1$\sigma$ and 3$\sigma$ uncertainty on the mean and dispersion), accounting for the upper/lower limits, and computed as described in Section \ref{subsec:distributions}. In the top panel we also show, when available, the mean of the distribution of in-situ clumps (gray line) and ex-situ satellites (pink line) found by \citet{Mandelker2014} in their cosmological simulations.}
   \label{fig:distr_meas}
\end{figure*}

\begin{table*}
\centering
\caption{Mean and dispersion of the distributions of observed and derived physical parameters for blobs and disks.}
\label{tab:distr_values}
\begin{tabular}{p{2.5cm}ccccccccccccc}
\toprule
\midrule
Parameter & Mean$_\mathrm{c}$ & Dispersion$_\mathrm{c}$ & Mean$_\mathrm{e}$ & Dispersion$_\mathrm{e}$ & Mean$_\mathrm{d}$ & Dispersion$_\mathrm{d}$ \\
(1)          &    (2)        & (3) & (4) & (5) & (6) & (7)    \\
\midrule
$\log($L$_\mathrm{cont})$                                     & 42.03 $\pm 0.11$ &  0.67 $\pm 0.10$ & 42.93 $\pm 0.16$ &  0.81 $\pm 0.14$ & 43.35 $\pm 0.11$ &   0.74 $\pm 0.09$ \\
$\log($L$_\mathrm{cont}/$L$_\mathrm{cont,gal})$   & -1.61 $\pm 0.06$ & 0.43 $\pm 0.05$ &  -0.89 $\pm 0.07$ &  0.37 $\pm 0.05$ &  -0.09$\pm 0.02$ &   0.12 $\pm 0.01$ \\
$\log($L$_\mathrm{line})$                                      & 40.47 $\pm 0.27$ & 1.22 $\pm 0.25$ & 40.98 $\pm 0.18$ & 0.74 $\pm 0.18$ & 41.67 $\pm 0.13$ &   0.75 $\pm 0.11$ \\
$\log($L$_\mathrm{line}/$L$_\mathrm{line,gal})$     & -1.20 $\pm 0.06$ &  0.36 $\pm 0.05$ &  -1.14 $\pm 0.09$ &  0.43 $\pm 0.07$ &  -0.49 $\pm 0.10$ &  0.32 $\pm 0.08$ \\
$\log($EW$)$                                                       & 2.27 $\pm 0.27$ &  1.32 $\pm 0.40$ & 1.80 $\pm 0.15$ & 0.72 $\pm 0.14$ &  2.02 $\pm 0.08$ &  0.52 $\pm 0.07$ \\
$\log($M$_\star)$                                                  & 8.15 $\pm 0.12$ &  0.75 $\pm  0.10$ &  9.41 $\pm  0.20$ &  1.05 $\pm  0.17$ &  9.69 $\pm  0.08$ &  0.55 $\pm  0.06$ \\
$\log($M$_\star/$M$_\mathrm{gal})$                       & -1.83 $\pm 0.13$ &  0.78 $\pm  0.11$ &  -0.77 $\pm 0.15$ &  0.78 $\pm  0.14$ &  -0.06 $\pm 0.04$ &  0.21 $\pm  0.06$ \\
$\log($SFR$)$                                                      & -0.07 $\pm  0.30$ &  1.46 $\pm  0.29$ &  0.68 $\pm  0.20$ &  0.81 $\pm  0.20$ &  1.34 $\pm 0.19$ &  1.11 $\pm  0.15$ \\
$\log($SFR/SFR$_\mathrm{gal})$                            & -1.69 $\pm 0.20$ &  0.91 $\pm  0.19$ &  -1.51 $\pm 0.21$ &  0.87 $\pm  0.20$ &   -0.23 $\pm 0.11$ &  0.60 $\pm  0.10$ \\
$\log($sSFR$)$                                                    & -7.89 $\pm 0.31$ &  2.01 $\pm  0.38$ &  -8.89 $\pm 0.36$ &  1.51 $\pm  0.35$ &   -8.39 $\pm 0.18$ &  1.01 $\pm  0.16$ \\
$\log($Age$)$                                                     & 8.29 $\pm  0.52$ &  2.66 $\pm  0.88$ &  9.01 $\pm  0.23$ &  1.14 $\pm  0.22$ &  8.73 $\pm  0.13$ &  0.88 $\pm  0.12$ \\
$\log(\Delta_\mathrm{MS})$                                   & -0.22 $\pm 0.83$ &  3.28 $\pm 1.28$ &  -0.55 $\pm   0.40$ &  1.54 $\pm  0.38$ &  -0.03 $\pm      0.19$ &  1.07 $\pm 0.17$ \\
$12 + \log($O/H$)$                                            & 8.56 $ \pm 0.10$ &  0.29 $\pm  0.11$ &  8.74 $\pm  0.11$  & 0.21 $\pm  0.14 $ & 8.54 $\pm  0.15$ &  0.49 $\pm  0.13$ \\
\bottomrule
\end{tabular}
\begin{minipage}{14cm}
Column: (1) Physical properties: continuum luminosity (L$_\mathrm{cont}$, in erg s$^{-1}$), continuum luminosity normalized by the total galaxy continuum luminosity (L$_\mathrm{cont}/$L$_\mathrm{cont,gal}$), emission line luminosity (L$_\mathrm{line}$, in erg s$^{-1}$), emission line luminosity normalized by the total galaxy emission line luminosity (L$_\mathrm{line}/$L$_\mathrm{line,gal}$), equivalent width (EW, in \AA), stellar mass (M$_\star$, in M$_\odot$), stellar mass normalized by the total galaxy stellar mass (M$_\star/$M$_\mathrm{gal}$), star formation rate (SFR, in M$_\odot$ yr$^{-1}$), star formation rate normalized by the total galaxy star formation rate (SFR/SFR$_\mathrm{gal}$), specific star formation rate (sSFR, in yr$^{-1}$), age (in yr), distance from the main-sequence ($\Delta_\mathrm{MS}$), gas-phase metallicity (12 + $\log($O/H$)$); (2) mean of the compact blobs' distribution; (3) dispersion (1$\sigma$, in dex) of the compact blobs' distribution; (4) mean of the extended blobs' distribution; (5) dispersion (1$\sigma$, in dex) of the extended blobs' distribution; (6) mean of the disks' distribution; (7) dispersion (1$\sigma$, in dex) of the disks' distribution.
\end{minipage}
\end{table*}

\subsection{Fitting the distributions of blob properties}
\label{subsec:distributions}

The non-detections in our sample (translating into lower- and upper-limits of the various physical properties) prevented us to directly determine and compare the distributions of the properties of compact and extended blobs. We therefore assumed that the physical properties of blobs and disks have log-normal distributions and inferred their parameters (i.e. mean $\mu$ and standard deviation $\sigma$) with a similar procedure to that adopted by \citet[who compare the SFR of AGN and main-sequence galaxies]{Mullaney2015}, \citet[who infer the AGN IR luminosity distribution]{Shao2010}, and \citet[who infer the SFR distribution of AGN hosts]{Bernhard2018}. The choice of a log-normal distribution is arbitrary and it might not represent the real distribution of the physical parameters of the blobs. However, simulations (\citealt{Mandelker2014}, \citealt{Tamburello2016}, \citealt{Tamburello2017}) suggest that the main properties of blobs (stellar mass, SFR, metallicity, age, and sSFR) are indeed log-normally distributed. Also the observational work of \cite{Guo2015} and \cite{Dessauges-Zavadsky2018} show that the luminosity and stellar mass functions of clumps are log-normally distributed. A log-normal stellar mass distribution for star clusters is also expected in the framework where star formation is driven by fragmentation induced by turbulent cascades. Since turbulence is a scale-free process, the stellar masses of star-forming regions are expected to have a log-normal distribution (\citealt{Elmegreen2006}, \citealt{Hopkins2013}, \citealt{Guszejnov2018}). 
Finally, assuming a log-normal distribution for the properties of the blobs is the most direct method to compare the observed properties of compact and extended components and to compare them with the models. Investigating whether other functional forms better describe these distributions goes beyond the scope of this paper and will need larger and deeper datasets.

We perform a maximum likelihood estimation. We maximize our likelihood function by randomly sampling the posterior distributions of the $\mu$ and $\sigma$ employing the affine invariant ensemble sampler of \cite{Goodman2010} fully implemented into EMCEE\footnote{EMCEE is publicly available at \texttt{http://dfm.io/emcee/current/ and we used the latest version}} \citep{Foreman-Mackey2013}. This method allows us to take into account uncertainties, upper- and lower-limits of the fitted parameters, and the resulting posterior probability distribution provides parameter uncertainties. We refer to \cite{Mullaney2015}, \cite{Bernhard2018}, and \cite{Scholtz2018} for more details on the method.

The results are reported in Figure \ref{fig:distr_meas} and Table \ref{tab:distr_values}, and we discuss them in Section \ref{sec:results}. We show the distribution of the physical parameters (stellar mass, SFR, sSFR, age, distance from the main-sequence, and metallicity) including measurements, upper- and lower-limits. We also show the mean and standard deviation of the distributions estimated through the approach described above. Finally, in Figure \ref{fig:distr_obs} we report the observables (continuum and emission line flux, equivalent width) used to derive the physical properties. Considering directly the observables instead of the derived physical properties for our analysis would bring to the same conclusions.

\section{Results and discussion}
\label{sec:results}

In this Section we discuss the distribution of the physical properties of blobs and we compare them with the properties of the underlying galaxy disks. Based on the observed properties of the blobs and the comparison with high-resolution cosmological simulations (\citealt{Mandelker2014}; \citealt{Mandelker2017}), we will discuss the nature and origin of these components.

\subsection{Properties of blobs}
\label{subsec:properties_blobs}

In our sample of 53 galaxies at $z \sim 1$ -- 3 we found 98 ``blobs'', namely components that depart from the diffuse stellar disk (Section \ref{sec:clumps_detection}). Among them, $\sim 70$\% are unresolved at the resolution of our data (PSF FWHM $\sim 0.15$'' $\sim 1.3$ kpc), whereas the remaining 30\% have extended S\'ersic profiles (Section \ref{subsec:light_profile}, Figure \ref{fig:flawchart_measuring}). To investigate the properties and nature of these components, we keep compact and extended blobs separated and we compare their physical properties with those of the underlying galaxy disk. In Figure \ref{fig:distr_meas} and Appendix \ref{app:observed_properties} (Figure \ref{fig:distr_obs}) we show the distribution of the physical properties of blobs and disks. We report the mean and standard deviation of the distributions in Table \ref{tab:distr_values}.

Blobs typically have lower integrated continuum luminosity than the underlying disks. This can be interpreted with blobs having lower stellar mass than the disks. This is mainly driven by the compact blobs, that have stellar masses in the range $7 \lesssim \log (\mathrm{M_\star/M_\odot}) \lesssim 9.5$, with a mean of $\log (\mathrm{M_\star/M_\odot}) \sim 8.15$, 1.5 dex lower than the underlying disks. The extended blobs instead have a broader distribution ($8 \lesssim \log (\mathrm{M_\star/M_\odot}) \lesssim 11$) peaking at $\log (\mathrm{M_\star/M_\odot}) \sim 9.4$. Their average mass is comparable to that of the disks and 1.3 dex more massive than that of the compact ones. Individual compact blobs enclose on average $\sim 10$\% of the stellar mass of the host disk, and $\sim 20$\% when summing the contribution of all the blobs belonging to a given galaxy. Extended blobs instead can have stellar masses comparable to that of the host galaxy.

Blobs typically have lower emission line luminosity and therefore star formation rate than the underlying disk, although their distribution is quite broad, with a scatter of $\sim$ 0.8 dex. Compact and extended blobs in our sample have comparable SFR, and the distribution peaks at SFR $\sim 1$ M$_\odot$ yr$^{-1}$ ($\sim 5$ M$_\odot$ yr$^{-1}$) for the compact (extended) components. Individual blobs make on average $\sim 10$\% of the total SFR of the host galaxy. When summing the contribution of all the compact blobs belonging to a given galaxy, they enclose on average $\sim 30$\% of the total star formation rate of the host disk. Extended blobs instead enclose on average $\sim 20\%$ of the total star formation rate of the host.

Blobs have on average a similar equivalent width as the underlying disks, although there is an evidence that the equivalent width distributions of the compact and extended blobs are different. Compact blobs have $\sim$ 0.5 dex higher equivalent width than the extended ones and their distribution shows a tail toward extremely high equivalent width ($\gtrsim 10^3$ \AA, see Figure \ref{fig:distr_obs}), a feature that is absent in the distribution of the extended blobs. The uncertainties on the mean equivalent width of the two populations are quite large though, due to the large number of upper and lower limits in our sample, and deeper observations of spatially resolved emission line and continuum maps are needed to draw stronger conclusions. The evidence of a higher equivalent width for compact blobs can be interpreted as compact blobs having an enhanced sSFR and/or younger ages with respect to the extended ones and the underlying disks. The sSFR and age distributions of compact blobs are broader ($\sigma \sim 2$ dex) than the extended ones ($\sigma \sim 1.5$ dex) and show a tail of enhanced sSFR and/or very young age ($\lesssim 10$ Myr) that is completely absent in the distributions of extended blobs. 

We also investigated the typical distance from the galaxy main-sequence of blobs and disks in our sample ($\Delta_{\mathrm{MS}} = \log($sSFR/sSFR$_\mathrm{MS})$, where sSFR$_\mathrm{MS}$ is the sSFR of a main-sequence source, with a given redshift and stellar mass). We find that blobs have a broad distribution around $\Delta_{\mathrm{MS}} \sim $ 0, consistent with that of the underlying disks, and therefore they seem on average to have a main-sequence star-forming mode. The distribution of compact blobs though shows a tail toward higher distances from the main-sequence ($\Delta_{\mathrm{MS}} \gtrsim 1$), with $\gtrsim 25$\% of the compact blobs having an enhanced sSFR with respect to the underlying disk and forming stars in starburst-like mode. The extended blobs instead have an average $\Delta_{\mathrm{MS}} \sim -0.5$, and typically have consistent or lower sSFR than main-sequence galaxies. Also in this case we have a large fraction of lower and upper limits and deeper data will be needed to confirm these findings.

Finally, for the subsample of blobs and galaxy disks with detected [OIII] and [OII] emission lines, we could compute the gas phase metallicity (Section \ref{subsec:metallicity}). Compact blobs show comparable metallicities as the underlying disks (12 + log(O/H) $\sim$ 8.5). The most massive extended blobs have comparable metallicity as the disks, whereas for the lower mass ones we could only estimate metallicity upper limits.

\begin{figure*}
	\includegraphics[width=\textwidth]{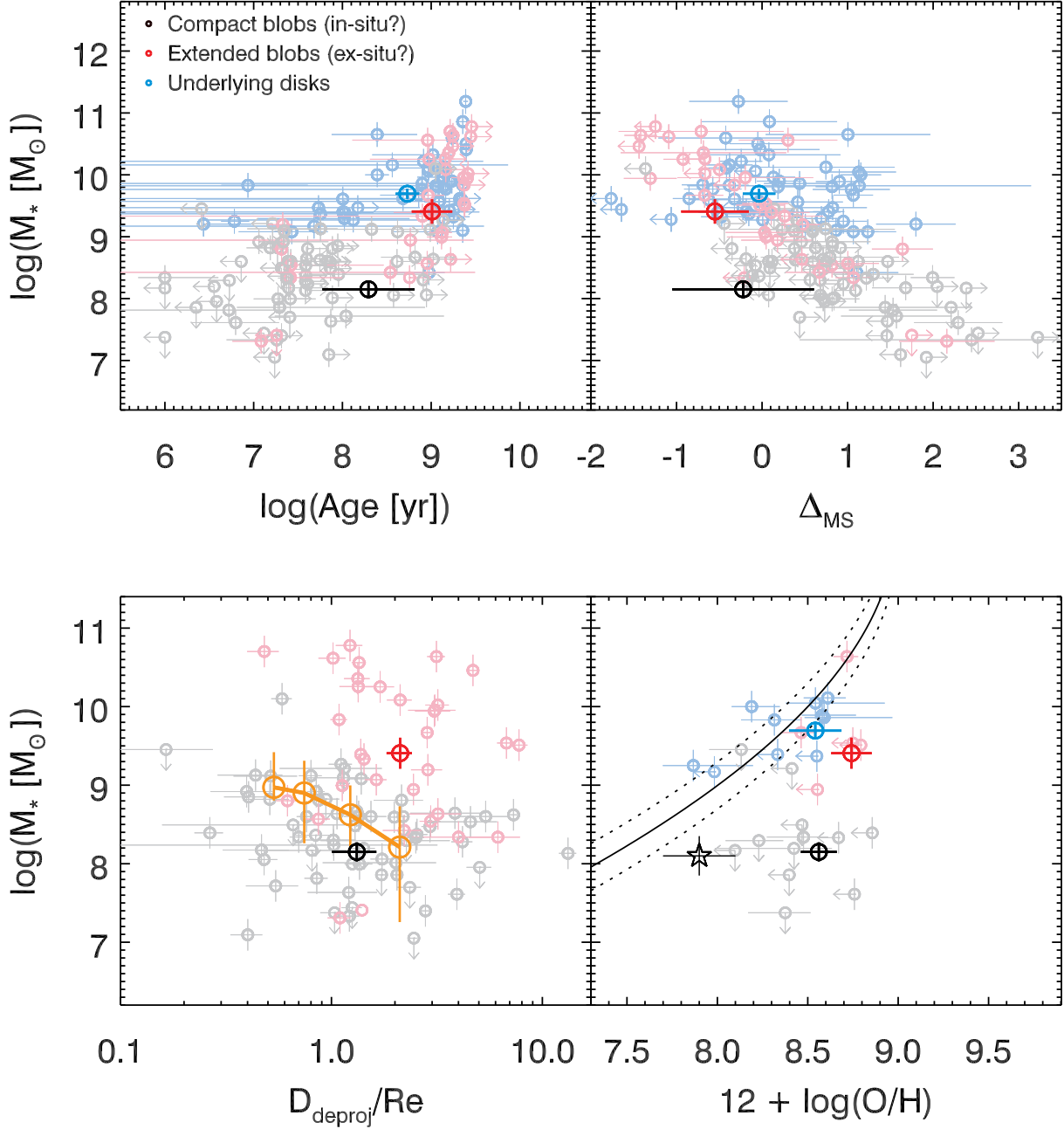}
  \caption{Physical properties of blobs and disks. In each panel, we show the relation between stellar mass and stellar age (\textit{top left panel}), distance from the main-sequence of star-forming galaxies (\textit{top right panel}), deprojected distance from the galaxy barycenter (\textit{bottom left panel}), and gas-phase metallicity (\textit{bottom right panel}). In the background we show individual compact (black symbols) and extend (red  symbols) blobs, and the underlying disks (cyan symbols), whereas the foreground symbols show the mean properties of each population. The stellar mass of clumps in bins of galactocentric distance from \citet{Guo2018} are shown in the bottom left panel (yellow symbols connected by the solid line). The mass -- metallicity relation of star-forming galaxies computed at the median redshift of the sample \citep{Maiolino2008} is shown in the bottom right panel together with the average location of dwarf galaxies with M$_\star \sim 10^8$ M$_\odot$ (\citealt{Calabro2017}, black star).}
   \label{fig:trends}
\end{figure*}

\subsection{In-situ clumps and ex-situ satellites}
\label{subsec:clumps_satellites}

We have considered the distribution of the physical properties of blobs and disks in our sample and estimated their mean and scatter (Section \ref{subsec:properties_blobs}). In the following we use these results to investigate the nature and origin of compact and extended blobs.

The compact and extended blobs in our sample appear to have different properties. The compact ones have on average fainter continuum and higher equivalent width than the extended components. This can be interpreted as the compact blobs having smaller masses and younger ages than the extended ones so, when looking at the stellar mass -- stellar age plane, the compact and extended blobs occupy distinct parts of the parameter space (Figure \ref{fig:trends}, top left panel). The extended blobs have comparable stellar masses and ages as the galaxy disks, they have a median effective radius of $\sim$ 2 kpc, S\'ersic index $n = 1.1$ (typical of disk galaxies), a comparable stellar mass surface density as disks ($\sim 2.5 \times 10^{8}$ M$_\odot$ kpc$^{-2}$), and are commonly found at larger distances from the galaxy barycenter than the compact blobs (Figure \ref{fig:trends}, bottom left panel). Finally, the extended blobs, as the disks, have a star formation mode that is consistent with that of main-sequence galaxies (Figure \ref{fig:trends}, top right panel). All this suggests that the extended blobs are satellites currently merging with the host galaxy. When considering the mass ratio between the primary galaxy and its satellites, we conclude that $\sim$11\% ($\sim$19\%) of our sample is undergoing major (minor) mergers (see Section \ref{subsec:merger_rate} for more detailes). The major mergers that we find based on the mass ratio of satellites and primary galaxy are also classified as such when using an independent non-parametric morphological classification performed on resolved stellar mass maps based on structural indices such as the galaxy asymmetry and M$_\mathrm{20}$ (\citealt{Cibinel2015} and Appendix \ref{app:mass_maps}).

The compact blobs instead likely have a different origin. The fact that compact blobs are unresolved even at the \textit{HST} resolution, that they are found at $\sim$ 1 kpc distance from the galaxy barycenter, that they have relatively small stellar masses ($\lesssim 15$\% of the underlying disk), but are actively forming stars suggests that they are star-forming regions likely originated due to disk instability and fragmentation of the galaxy disk (\citealt{Bournaud2014}; \citealt{Mandelker2017}). The in-situ formation of the compact blobs is further supported by their metallicity. In fact, while the disk properties are consistent with the stellar mass -- metallicity relation of $z \sim 2$ star-forming galaxies (e.g. \citealt{Maiolino2008}; \citealt{Zahid2014}), compact blobs instead show metallicities inconsistent with the mass-metallicity relation (Figure \ref{fig:trends}, bottom right panel). They have a comparable metallicity to the disks, but $\sim$ 1.5 dex lower stellar masses, so they are $\sim$ 1 dex above the mass -- metallicity relation. Metallicity measurements for statistical sample of galaxies with M$_\star \sim 10^8$ M$_\odot$ at $z \sim 2$ are still lacking and therefore at the low-mass end we are showing an extrapolation of the mass-metallicity relation derived for galaxies with M$_\star \gtrsim 10^9$ M$_\odot$. In Figure \ref{fig:trends} we also show the average location of dwarf galaxies with M$_\star \sim 10^8$ M$_\odot$ at $z \lesssim 1$ (\citealt{Kirby2013}, \citealt{Calabro2017}, \citealt{Hidalgo2017}). Despite some of them may have gas-phase metallicities up to 12 + $\log(\mathrm{O/H}) \sim 8.5$ \citep{Sanchez-Almeida2018}, on average our  
compact blobs seem to be $\sim 0.5$ dex more metal-rich than dwarf galaxies (Figure \ref{fig:trends}). The high metallicity of compact clumps further suggests that they formed in-situ, due to the gravitational collapse of pre-enriched gas in unstable regions of the galaxy disk. The young ages of the blobs reported in Figure \ref{fig:trends} support these conclusions. In fact, the metallicity of star-forming regions is altered in about one galactic dynamical time ($\gtrsim$ 100 Myr), increasing due to the active star formation and internal production of metals. The fact that our sample clumps with metallicity measurements have ages $\lesssim$ 50 Myr points toward the conclusion that they formed in-situ from metal-rich gas, since this timescale is too short for the gas to be self-enriched due to internal star formation \citep{Bournaud2016}.  Finally, the metallicity of our sample of clumps does not clearly correlate with their star formation rate, as indeed expected if they formed from pre-enriched material, although larger statistical samples are needed to confirm this finding.

Our interpretation of extended blobs being accreting satellites and compact components being in-situ star-forming clumps is also supported by simulation results. In their simulations, \cite{Mandelker2014} study clumps in $z \sim 1$ -- 3 galaxies to understand their origin and fate. In particular, they identify blobs in the gas density maps and divide them into two populations, based on their dark matter content: those that are formed in-situ due to Toomre instability and do not contain dark matter, and the satellites that are formed ex-situ and are embedded into their own dark matter halo. In Figure \ref{fig:distr_meas} we compare our observations with the results of simulations. The models by \cite{Mandelker2014} show that the satellites have larger masses, older ages, and lower specific star formation rates than the in-situ clumps, but both populations have near-solar metallicities. This is consistent with the physical properties of our extended and compact blobs, supporting the scenario in which the extended blobs in our sample are typically accreting satellites, whereas the compact ones are in-situ formed clumps.

The peak of the simulated distributions are in good agreement with our observations. Small differences between observed and simulated distributions (e.g. the simulated blobs have slithgly smaller masses compared to their host galaxy than the ones we estimate) can be attributed to the large scatter (up to 1.5 dex in the case of the satellites, Figure \ref{fig:distr_meas}). Alternative causes might be due to the limiting flux of our observations, preventing the detection of blobs with masses $\lesssim 10^7$ M$_\odot$ and/or the limited resolution of our observations ($\sim 1$ kpc) that does not allow us to deblend small clustered blobs \citep{Behrendt2016} and some massive blobs might therefore be the result of less massive, blurred ones. Finally, the differences might be due to the stellar feedback model adopted in the simulations that could have an important impact on the formation and survival of low-mass clumps \citep{Mandelker2017}. 

\begin{figure}
	\includegraphics[width=0.5\textwidth]{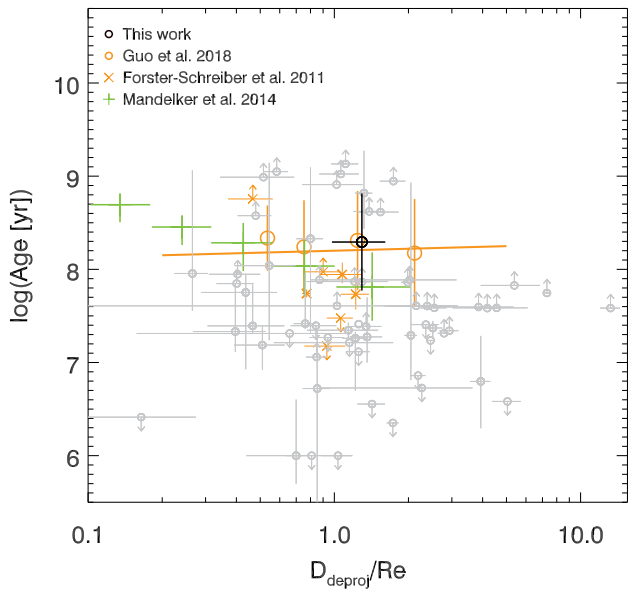}
  \caption{Age of the clumps (i.e. compact blobs) as a function of their deprojected distance from the galaxy barycenter normalized by the galaxy effective radius. We report measurements for individual clumps (gray symbols) and their average properties (black thick circle). Clumps' age in bins of galactocentric distance from \citet{Guo2018} at $z \sim 2$ are shown (yellow circles) together with the age -- distance flat trend that they find (yellow line). The seven clumps found in the star-forming galaxy BX482 at $z = 2.26$ by \citet{ForsterSchreiber2011} are also shown (yellow crosses, with arrows indicating upper/lower limits). Finally, we report the age -- distance trend predicted by numerical simulations for in-situ formed clumps at $z \sim 2$ \citep[green crosses, with error bars]{Mandelker2017}.}
   \label{fig:age_gradient}
\end{figure}

\subsection{Constraining clumps' lifetime}
\label{subsec:clumps_lifetime}

It is currently debated whether clumps are short- or long-lived. In fact simulations considering various recipes for stellar feedback predict different scenarios: clumps could be rapidly disrupted by the intense stellar feedback on timescales of $\sim$ 50 Myr (e.g. \citealt{Genel2012}; \citealt{Tamburello2016}; \citealt{Oklopcic2017}) or, if they transform their gas into stars quickly enough, they could survive stellar feedback and have lifetimes of $\sim 500$ Myr (e.g. \citealt{Ceverino2012}; \citealt{Bournaud2014}, \citealt{Mandelker2014}, \citealt{Mandelker2017}). In this section we will constrain the clump formation rate and lifetime in a statistical way, following a procedure similar to the one adopted in \cite{Zanella2015}. For this calculation we only considered the compact blobs in our sample, that likely formed in-situ.

The clump formation rate indicates the typical number of clumps that are formed in a galaxy per Gyr. To estimate it we considered the number of young clumps in our sample, namely the number of clumps with equivalent width larger than the one expected for a stellar population of 20 Myr \citep{Zanella2015}, as predicted by stellar population synthesis models. The clump formation rate can be computed as CFR = N$_\mathrm{young}/(t_\mathrm{V} \mathrm{N_\mathrm{gal}})$. $t_\mathrm{V}$ is the ``visibility window'', namely the period of time during which a clump can be considered ``young''. In our case, we set it to 20 Myr, as simulations predict that this is the timescale during which young clumps are starbursting and later they keep forming stars in a more quiet mode (\citealt{Bournaud2014}, \citealt{Zanella2015}). N$_\mathrm{young}$ is the number of young clumps in our sample, and N$_\mathrm{gal}$ is the total number of galaxies in our sample, given by the galaxies in our survey plus their merging satellites. We include the merging satellites in this calculation as they could also be clumpy. In Figure \ref{app:fig_stat} there are indeed some examples of extended blobs that seem to host more compact ones (e.g. ID123, ID799), although they were not found by our pipeline, likely due to the low significance. Properly investigating the clumpiness of satellites goes beyond the scope of this paper though and deeper data might be needed.  If we were not to include the satellites in our CFR and lifetime calculation, our estimate of the CFR (lifetime) would increase (decrease) by $\sim 30$\%. The clump formation rate is expected to depend on the stellar mass of the clumps, with low mass clumps being potentially more frequent than massive ones (e.g. \citealt{Mandelker2014}, \citealt{Tamburello2016}, \citealt{Dessauges-Zavadsky2018}, \citealt{Guo2018}). Therefore, we divided our sample of young clumps in two mass bins, using a stellar mass M$_\star = 10^9$
 M$_\odot$ as dividing threshold. For the low mass clumps (M$_\star \leq 10^9$ M$_\odot$) we obtained a CFR = 9.7 Gyr$^{-1}$, whereas for the high mass clumps (M$_\star > 10^9$ M$_\odot$) the clump formation rate is CFR = 1.7 Gyr$^{-1}$.

We then estimated the lifetime of low- and high-mass clumps as follows. Assuming that the stellar mass of clumps increases as they get older, we estimated from our own sample and from the literature (\citealt{ForsterSchreiber2011}, \citealt{Guo2018}) the average number of clumps per galaxy (N$_{\mathrm{old/gal}}$) with old ages (age $>$ 300 Myr) and a stellar mass larger than the one observed for our young clumps (on average M$_\star \sim 3\times10^8$ M$_\odot$ and M$_\star \sim 1.5\times10^9$ M$_\odot$ in the low- and high-mass bins respectively). We therefore statistically estimated the clump lifetime as LT = N$_{\mathrm{old/gal}}$/CFR. We found that low-mass clumps have typical lifetimes of $145^{+293}_{-27}$ Myr, whereas the high-mass ones have lifetimes of $650^{+1770}_{-341}$ Myr. We estimated the uncertainties considering the Poisson error associated to the number of young clumps in our sample and to the number of clumps per galaxy taken from the literature. Some simulations predict that more massive clumps form more frequently in more massive galaxies (e.g. \citealt{Dekel2009}, \citealt{Ceverino2012}). If we were to calculate the lifetime of massive clumps considering only the galaxies in our sample with stellar mass M$_\star \gtrsim 5\times 10^9$ M$_\odot$ (so that the clump would enclose $\sim 20$\% of the galaxy stellar mass), we would obtain a lifetime $\sim 320^{+1140}_{-160}$ Myr. This is consistent with the estimate obtained considering all galaxies in our sample and the conclusions would not change.

Low-mass clumps seem to be formed more frequently than high-mass ones and have on average 4.5 times shorter lifetimes. This could be due to the fact that they are more easily disrupted by stellar feedback and/or by the fact that during their lifetime they merge and form higher-mass clumps \citep{Mandelker2017}. These results favour simulations predicting that clumps are long-lived (lifetime $\gtrsim$ 100 Myr) and potentially play a relevant role in the mass assembly of the host galaxy.

We highlight though that the spatial resolution of our observations is $\sim$ 0.1'' -- 0.2'' (depending on the \textit{HST} band), corresponding to $\gtrsim 1$ kpc at $z \sim 2$. Some of the blobs in our sample (especially the compact ones) might be the result of smaller clumps blended due to the lack of resolution and appearing as a kpc-size star-forming region. This might bias our size and mass measurements toward larger and more massive clumps than they are in reality and potentially affect our statistical estimate of their lifetime. Observations with better resolution will be needed to assess this caveat.

\subsection{Constraining clumps' migration}
\label{subsec:clumps_migration}

Simulations predicting that clumps are long-lived and survive stellar feedback for $\sim$ 500 Myr also expect them to migrate inward (e.g. \citealt{Ceverino2010}, \citealt{Bournaud2014}, \citealt{Mandelker2014}, \citealt{Mandelker2017}). They find massive clumps to undergo dynamical friction with the underlying galaxy disk, dissipate kinetic energy and angular momentum, and spirale towards the center of the galaxy potential well. In addition, since clumps lie mostly in the disk plane, they also undergo gravity torques from the neighbouring regions of the disk. This accelerates their inward migration and funnels additional gas toward the center. These simulations show that the coalescence of clumps and gas contributes to the formation of the galaxy bulge. The inward clump migration is expected to leave, as observational signature, an age gradient, with older clumps found on average at smaller galaxy radii (i.e. closer to the galaxy barycenter). Statistical samples of clumps are needed to verify this scenario. 

We investigated whether a gradient of clumps' age with galactocentric distance was found in our sample. In Figure \ref{fig:age_gradient} we show our sample of clumps (only the compact blobs) and we compare it with literature clump samples at similar redshift (\citealt{ForsterSchreiber2011}, \citealt{Guo2018}) and the predictions of simulations \citep{Mandelker2017}. Our sample has a mean age of $\sim$ 100 Myr (consistent with the typical age of the clumps found by \citealt{Guo2018}) and no strong age gradients are detected. This is consistent with the results by \cite{Guo2018} at similar redshift ($z \sim 2$). \cite{ForsterSchreiber2011} argued for a possible age gradient, but their result was likely driven by the small number statistics, and their measurements are in good agreement with our findings. In Figure \ref{fig:age_gradient} we also report the average trend expected from simulations \citep{Mandelker2017}. This seems to be mostly driven by the innermost clumps with galactocentric distance $< 0.5$ R$_\mathrm{e}$, a part of the parameter space that is poorly sampled by the observations due to the limited spatial resolution ($\sim$ 1 kpc at $z \sim 2$). In our sample we have only one clump with distance $< 0.2$ R$_\mathrm{e}$. It has very young age, but its age measurement is highly uncertain due to the proximity with the central bulge where the deblending is particularly difficult due to the poor spatial resolution. Higher resolution observations will be needed to properly find clumps close to the galaxy center, deblend them, and measure their properties (e.g. distance, age) to assess whether observations at this redshift are in agreement with numerical simulations.

At lower redshift ($z \sim 0.5 - 1$) a steep gradient of clumps' age with galactocentric distance was found by \cite{Guo2018} and it was interpreted as a signature of clumps inward migration. Possible reasons for the different results obtained at $z \sim 2$ and at lower redshift are the following: the signature of migration is now yet visible at $z \sim 2$ as the timescale for clumps migration is longer; at lower redshift the better spatial resolution of the observations allows to probe the innermost regions of the disk and better constrain the properties of clumps closer to the galaxy center; measurement uncertainties, especially in the age determination, hide a possible gradient. Larger samples of clumps observed in different redshift ranges and with better spatial resolution will help to clarify whether and how clumps migrate.

\subsection{Constraining the merger fraction}
\label{subsec:merger_rate}

In this Section we focus on the galaxies with satellites (i.e. extended blobs). We find that 32\% of our sample galaxies have at least one nearby satellite and 13\% of them have multiple satellites.  The average stellar mass ratio of the satellites is 1:5, but it has quite a large spread, ranging from 1:10 to 1:1. If we consider the mass ratio 1:4 as criterion to separate minor and major mergers (e.g. \citealt{Lotz2011}), $\sim$17\% (23\%) of our sample galaxies are undergoing a major (minor) merger. Some of the satellites might be chance alignments, as some of them are only detected in the continuum and not in the emission line maps (so a secure measurement of their redshift is missing). However, if we only consider the galaxies and satellites with secure redshift, the merger fractions decrease to 25\% for galaxies with at least one satellite and 11\% for galaxies with multiple satellites. Among them, 11\% (19\%) are major (minor) mergers. Likely the correct merger fractions are in between these two extreme estimates (i.e. only some of the satellites in our sample are chance alignments). These results are in broad agreement with the findings by \cite{Man2016} that analyze a large statistical sample of galaxies and find a fraction of major (minor) close pairs of $\sim$ 10\% ($\sim$ 15\%) at $z \sim 1.5 - 2.5$. Also the mean stellar mass ratio of our mergers (1:5) is comparable to that found by \citet[number-weighted mean stellar mass 1:4 -- 1:5]{Man2016}. Also \cite{Cibinel2018} investigates the close pair fraction for a sample of main-sequence galaxies, separating clumpy disks from galaxy major mergers (stellar mass ratios 1:1 -- 1:6) by using non-parametric morphological classifications of the spatially-resolved stellar mass maps. They also find a merger fraction of $\sim$ 10\% -- 15\% at redshift $z > 1$, consistent with ours. Previous merger fraction measurements based on pair statistics (\citealt{Williams2011}, \citealt{Newman2012}, \citealt{Mantha2018}) are in good agreement with our findings. We stress however that all these samples are not fully homogeneous in terms of galaxy separation, stellar mass, and stellar mass ratio and matching these samples with ours goes beyond the scope of this paper. We are referring to the literature not to have a fully consistent comparison, but rather to put our work into context. Finally, we are not comparing to samples of morphologically-classified mergers (e.g. \citealt{Lotz2008}, \citealt{Conselice2009}, \citealt{Bluck2012}) as the the selection is too different with respect to the one that we have adopted in this work (i.e. we did not classify mergers based on their disturbed morphology). We also highlight that our sample of mergers have a projected distance d$_{\mathrm{proj}} < 10$ kpc, so they are much closer pairs than the ones usually studied in the literature (d$_\mathrm{proj} \sim$ 10 -- 30 kpc h$^{-1}$ in \citealt{Williams2011}, \citealt{Newman2012}, \citealt{Man2016}, d$_\mathrm{proj} \leq$ 30 kpc in \citealt{Cibinel2018}, and d$_\mathrm{proj} \sim$ 5 -- 50 kpc in \citealt{Mantha2018}).

To investigate whether mergers increase the disk instability and clump formation, we compared the observed number of clumps (i.e. compact blobs) per galaxy in isolated and merging galaxies. We find that when a satellite is present, the average number of clumps per galaxy is $1.2 \pm 0.2$ (where the uncertainty has been computed considering Poissonian uncertainties). In case no satellites are detected, $2.0 \pm 0.2$ clumps per galaxy are found instead. Taking our results at face value, it seems that during mergers the number of clumps in the galaxy decreases. We also do not find a preferred spatial distribution of clumps across the galaxy disk with respect to the location of the impacting satellite. However larger samples of galaxies and satellites with secure spectroscopic redshift will be needed to validate these results.

\subsection{Implications for the mass assembly of disk galaxies}
\label{subsec:assembly}

In our sample of star-forming galaxies at $z \sim 1$ -- 3, we find that 70\% host additional components (``blobs'') on top of a diffuse disk and among the blobs, 30\% are likely accreting satellites and 70\% are clumps formed in-situ due to gravitational instability. These number densities are in remarkable agreement with those from simulations \citep{Mandelker2014}, finding that $\sim 75$\% of the blobs hosted by $z \sim 1$ -- 3 galaxies are formed in-situ due to disk instability and the remaining $\sim 25$\% are merger remnants. We divided our sample of blobs based on their spatial profile (compact PSF-like blobs versus extended ones with S\'ersic profile), we found two populations with different physical properties on average, and we concluded that statistically the compact blobs are in-situ formed clumps and the extended ones are accreting satellites. However there might be some contamination: part of the compact blobs might be small satellites (especially those found at large distances from the galaxy barycenter) and some of the extended blobs might be large clumps or complexes of multiple nearby blurred clumps or, potentially, other galaxy structures (e.g. (proto-)spiral arms, see in particular ID507, ID508, ID580 in Figure \ref{app:fig_stat}). Data with better spatial resolution (e.g. \textit{JWST}/NIRCam, ELT/HARMONI) will be needed to disentangle among these different scenarios. We checked that by dividing our sample of blobs based on their continuum luminosity and distance from the galaxy barycenter instead (i.e. without considering their spatial profile) we would obtain consistent conclusions, with the most distant and luminous blobs being on average more massive and older than the ones that are less luminous and closer to the galaxy center. This supports the fact that our sample is mainly made of two populations with different properties (i.e. clumps and satellites). We also highlight that our study is by construction biased toward galaxies with bright emission lines. In fact to astrometrically calibrate emission line maps we cross-correlated spectra with three different grism orientation (Section \ref{subsec:emission_line_maps}). For this procedure to work each sample galaxy needs to be detected in the individual emission line maps (S/N $\gtrsim$ 3). This bias however does not seem to affect our results, as the distribution of galaxy properties (e.g. star formation rate, stellar mass, effective radius) seems to be consistent with that of typical $z \sim 1$ -- 3 star-forming galaxies (Figure \ref{fig:main_sequence} and \ref{fig:mass_size_relation}) and the properties of our in-situ clumps and ex-situ satellites are consistent with those expected from simulations.
Satellites make up to $\lesssim$80\% of the stellar mass and $\sim$ 20\% of the star formation rate of the host galaxy, whereas clumps have a smaller contribution (on average $\sim$20\% and $\sim$ 30\% to the stellar mass and star formation rate respectively).

The fact that our sample clumps seem to be long-lived indirectly implies that they likely play an important role in growing the bulge of galaxies. In fact, according to simulations (\citealt{Elmegreen2008}, \citealt{Ceverino2010}, \citealt{Genel2012}, \citealt{Bournaud2014}, \citealt{Mandelker2014}), if clumps are long-lived they can induce gravitational torques in the disk that bring gas inward and contribute to bulge formation. Therefore direct clumps migration is not strictly required to grow a bulge, and the fact that they survive stellar feedback for $\gtrsim 100$ Myr might be sufficient for them to play an important role in galaxy evolution. Future works with larger samples of clumps observed in different redshift ranges and with better spatial resolution (e.g. with \textit{JWST} and the ELTs) will help to clarify whether and how clumps migrate.

Finally, based on the fact that young clumps seem to have comparable metallicity as the host disk and different stellar mass and age distribution with respect to the satellites, we have concluded that the clumps with age $\lesssim 300$ Myr ($\sim$ 80\% of the sample) have formed in-situ due to disk instability. The oldest clumps (age $\gtrsim 300$ Myr) have on average similar age as the accreting satellites and $\sim$ 0.5 -- 0.8 dex smaller stellar mass (Figure \ref{fig:trends}). Some of them might therefore be remnants of accreted satellites that have not been fully disrupted, but have been stripped during coalescence and appear as compact old star-forming regions. This is in agreement with the expectations from simulations (e.g. \citealt{Mandelker2014}, \citealt{Bournaud2016}) predicting that only a small fraction of clumps have an ex-situ origin. To confirm these findings it will be key to investigate whether these clumps show large deviations from the underlying velocity field of the host. Due to the low resolving power of our grism data ($R = 130$ at 1400 nm) we could not perform this test and follow-up spectroscopic observations with IFU instruments and better resolving power (e.g. VLT/ERIS, \textit{JWST}/NIRSpec, ELT/HARMONI with $R > 1000$) will be needed.

\section{Conclusions}
\label{sec:conclusions}

In this paper we discuss the contribution of giant-star forming clumps and accreting satellites to the build-up of stellar mass in redshift $z \sim 2$ galaxies and their role in galaxy evolution. We considered a sample of 53 star-forming galaxies at $z \sim 1$ -- 3 with \textit{HST} broadband (F140W, F105W, F606W) and slitless spectroscopic data. We simultaneously analyzed spatially resolved continuum and emission line maps (H$\alpha$, [OIII], H$\beta$, and/or [OII] depending on the redshift of the galaxy), modelling the galaxy disk with a S\'ersic profile and residual blobs with PSF or S\'ersic profiles. We initially fitted the blobs with a PSF profile, but found 30\% had significant residuals from these fits and were better described by a extended/resolved (S\'ersic) model. The median size of the extended blobs is 2 kpc and their S\'ersic index $n \sim 1.1$. We estimated the physical properties of blobs and underlying galaxy disks (intrinsic luminosity, star formation rate, stellar mass, stellar age, distance from the main-sequence, gas-phase metallicity). In the following we summarize the main conclusions we reached.

\begin{itemize}
\item Extended blobs and disks have similar properties: continuum fluxes, stellar masses, ages (or specific star formation rates), metallicity. Compact blobs instead seem to be a different population: they have lower continuum fluxes and stellar masses, and younger ages (or higher specific star formation rates) than the disks. However compact blobs have comparable metallicity to the disks, despite their $\sim 30$ times lower stellar mass (Figures \ref{fig:distr_meas} and \ref{fig:trends}).

\item Extended blobs make up to $\sim$80\% of the stellar mass and $\sim$ 20\% of the star formation rate of the host galaxy, whereas compact ones have a smaller contribution ($\sim$ 20\% and $\sim$ 30\% to the stellar mass and star formation rate respectively). 

\item The fact that extended satellites are found on average at larger distances from the galaxy barycenter, are larger, more massive, and older than the compact ones, and have similar physical properties as the disks suggests that the extended blobs are likely merging satellites whereas the compact blobs could be formed in-situ due to violent disk instability. The in-situ formation of compact blobs is also supported by the fact that they have a metallicity similar to the underlying disks. Due to their 1.5 dex lower stellar mass, if they were lying on the mass--metallicity relation, they would have been more metal-poor than the host galaxy (metallicity difference $\sim 0.9$ dex). This result can be explained if the compact blobs formed in-situ due to disk instability, from the pre-enriched gas of the galaxy disk.

\item We compared our observations with the predictions of the cosmological simulations by \cite{Mandelker2014} and \cite{Mandelker2017}. 
The physical properties (stellar mass, star formation rate, age, and metallicity) of simulated in-situ formed clumps and ex-situ satellites reasonably agree with those we observe for compact and extended blobs, respectively. 
This further supports the idea that the majority of the compact blobs observed in our sample are in-situ formed clumps, whereas the bulk of the extended ones are accreted satellites.

\item By considering only the in-situ formed clumps (i.e., the compact blobs) in our sample, we statistically estimated the clump formation rate and lifetime. We divided the sample in low- and high-mass clumps (dividing mass threshold $\sim 10^9$ M$_\odot$). We obtained a clump formation rate of $\sim$ 9.7 Gyr$^{-1}$ (1.7 Gyr$^{-1}$) and a lifetime of $\sim$ 145 Myr (650 Myr) for the low-mass (high-mass) clumps. The shorter lifetime found for the low-mass clumps could be due to the fact that they are more easily disrupted by stellar feedback and/or by the fact that during their lifetime they merge and form higher-mass clumps. These results support simulations predicting long-lived clumps.

\item We investigated whether clumps migrate towards the center of the galaxy and potentially contribute to bulge formation. We looked for an age gradient with distance from the galaxy barycenter (Figure \ref{fig:age_gradient}). We found a rather flat age distribution in the range of galactocentric distances that we explored, in broad agreement with the observational findings by \cite{Guo2018} and \cite{ForsterSchreiber2011} at similar redshift ($z \sim 2$), and with the models of \cite{Mandelker2014}.

\item We constrained the merger fraction by considering the satellites (i.e., the extended blobs) that we found in our sample. We concluded that $\sim 25$\% of our sample galaxies have a nearby (distance $\lesssim$ 10 R$_\mathrm{e}$) satellite and $\sim$ 11\% of them have multiple satellites. Among them, $\sim$ 11\% ($\sim$ 19\%) are major (minor) mergers, when considering a stellar mass ratio 1:4 to distinguish among major and minor mergers. The typical galaxy-satellite stellar mass ratio is 1:5 (but it ranges from 1:10 to 1:1). In galaxies undergoing a merger the number of clumps seems to be smaller (1.2 $\pm$ 0.2) than in isolated sources (2.0 $\pm$ 0.2), although this result is only marginally significant and larger samples are needed to confirm it.
\end{itemize}

\section*{Acknowledgements}

We thank the referee, D. Ceverino, for his comments and suggestions. AZ thanks B. Haeussler for helpful feedback about the GALFITM fitting algorithm and A. Dekel for interesting discussions. FV acknowledges the Villum Fonden research grant 13160 ``Gas to stars, stars to dust: tracing star formation across cosmic time'' and the Cosmic Dawn Center of Excellence funded by the Danish National Research Foundation. JSA acknowledges support from the Spanish Ministry of Economy and Competitiveness through grant AYA2016-79724-C4-2-P. MB acknowledges support from DFG BU/842/25-1.




\bibliographystyle{mnras}
\bibliography{bibliography} 


\newpage
\appendix

\section{Creating the emission-line maps}
\label{app:cross_correlation}

Our \textit{HST} G141 grism observations were executed along three position angles ($\sim$0, -30, +15 degrees), to correct each spectrum for contamination of nearby sources and allow for a proper astrometric calibration of the emission line maps. For each detected line, emission line maps were created by maximizing the cross-correlation between the spectral images with the three different grism orientations and the continuum probed by the F606W filter. We iteratively shifted the 2D grism spectral cutouts along the dispersion direction, computing the cross-correlation between them and the F606W image at each step. We fitted the curve obtained considering the cross-correlation as a function of the shift of each image: the maximum of the curve indicates the relative position of the images that gives the correct astrometric calibration. The uncertainty associated to our best solution has been computed by propagating the errors of the parameters of the fit. Once the relative position of the images that maximizes the cross-correlation has been found, the spectral maps were combined with WDRIZZLE \citep{Fruchter2002}, weighting each single orientation by its corresponding exposure time. The absolute astrometric calibration along the dispersion direction of the grism was determined from the cross-correlation of the [OIII] or H$\alpha$ spectral images (depending on the redshift of the source), as these are the lines detected with the highest signal-to-noise ratio (S/N). The astrometry of the H$\beta$ and [OII] emission maps was afterwards tied to that of the [OIII] or H$\alpha$ maps. Knowing the wavelength solution of each spectrum and the shift applied to obtain the correct astrometric calibration, we could associate a redshift to each emission line map ($z_{\mathrm{map}}$). We note that, for the same galaxy, maps with different position angles could have a different associated redshift. This is due to the fact that the shifts applied during the cross-correlation procedure could have been different, even if \textit{a priori} all the spectral images of the galaxy should have the same redshift, independently of their position angle. We could not force the redshift to be the same due to the presence of distorsions in the data. Furthermore, during the cross-correlation we only allowed the maps to shift along the dispersion direction, assuming that the spectra were perfectly aligned with the broad-band imaging along the cross-dispersion direction (e.g. the spectral trace of the continuum is aligned with the barycenter of the galaxy in the imaging). This was in general correct, although distorsions might sometimes prevent a good alignment. Due to the different position angles of each map, misalignments along one spatial direction would also partially affect the other. To quantify the effect of the distorsions, we proceeded as follows. We computed a mean redshift for each galaxy ($z_{\mathrm{aver}}$), averaging the $z_{\mathrm{map}}$ associated to emission line maps with different position angles. To estimate the distortions, for each position angle of the maps, we forced the reduced Chi square to be equal to 1:
\begin{equation}
\chi^2_{\mathrm{red}} = \frac{1}{N}\Sigma_{i = 0}^N \frac{(D_{\mathrm{i}} - D_{\mathrm{aver}})^2}{\epsilon_{\mathrm{z,i}}^2 + \epsilon_{\mathrm{z,aver}}^2 + \sigma_{\mathrm{D}}^2}
\end{equation}
where \textit{N} are the degrees of freedom (basically the number of our sample galaxies), $D_{\mathrm{i}}$ is the difference $z_{\mathrm{map}} - z_{\mathrm{aver}}$, and $D_{\mathrm{z,aver}}$ is the average of those differences. Besides, $\epsilon_{\mathrm{z,i}}$ and $\epsilon_{\mathrm{z,aver}}$ are the uncertainties associated to $D_{\mathrm{i}}$ and $D_{\mathrm{aver}}$ respectively. Finally, $\sigma_{\mathrm{D}}$ is an additional term needed to obtain $\chi_{\mathrm{red}}^2 = 1$: it accounts for the average distorsion affecting the maps with the considered position angle. We concluded that the distorsions are smaller than 0.06'', for all position angles, and repeating the analysis separately on the \oiii, \ha ~and \oii ~emission line maps, we checked that they do not depend on the wavelength. We also tried to divide the galaxies based on their location in the field of view, to check if the distorsions could depend on it. We concluded that there is no trend with the right ascension and declination of the galaxy and therefore we can consider that the distorsions in the dispersion and cross-dispersion directions of the spectral images are comparable.

A possibility to correct for these distorsions is to run iteratively our cross-correlation procedure allowing the spectra to shift along the dispersion direction first, then to fix the solution that has been found, and to repeat the cross-correlation allowing offsets along the cross-dispersion direction instead. This procedure would need to be repeated until convergence is reached. However, this method is extremely time consuming, therefore we decided to apply the cross-correlation procedure shifting the spectra along the dispersion direction only, and to refine \textit{a posteriori} potential misalignments affecting some galaxies. To this aim, in case multiple blobs in a single galaxy were found, we used them to refine the alignment between maps and imaging, computing the additional offset that was needed to better overlap the single blobs. We visually inspected all the shifted maps to check the reliability of the alignment. We found that the average shift of these maps is smaller than 0.03'', consistent with the distorsions that we estimated with the cross-correlation procedure. We note that these additional shifts were estimated using the \oiii ~or \ha ~emission line maps and then applied to the spectral images with lower S/N (\hb ~and \oii). Additionally, when fitting each galaxy light profile to estimate the blobs' flux, we allowed the best fit model to rigidly shift to correct for possible residual misalignments of the continuum and emission line maps (we discuss this in Section \ref{sec:clumps_detection}). 

A subsample of our galaxies was followed-up with longslit MOIRCS spectroscopy: we verified that the average redshift $z_{\mathrm{aver}}$ that we computed for each galaxy with the cross-correlation technique was in agreement with the one derived from higher resolution MOIRCS spectra \citep{Valentino2015}. 

\section{Our sample of galaxies and blobs}
\label{app:fig_stat}

\begin{figure*}
	\includegraphics[width=0.9\textwidth]{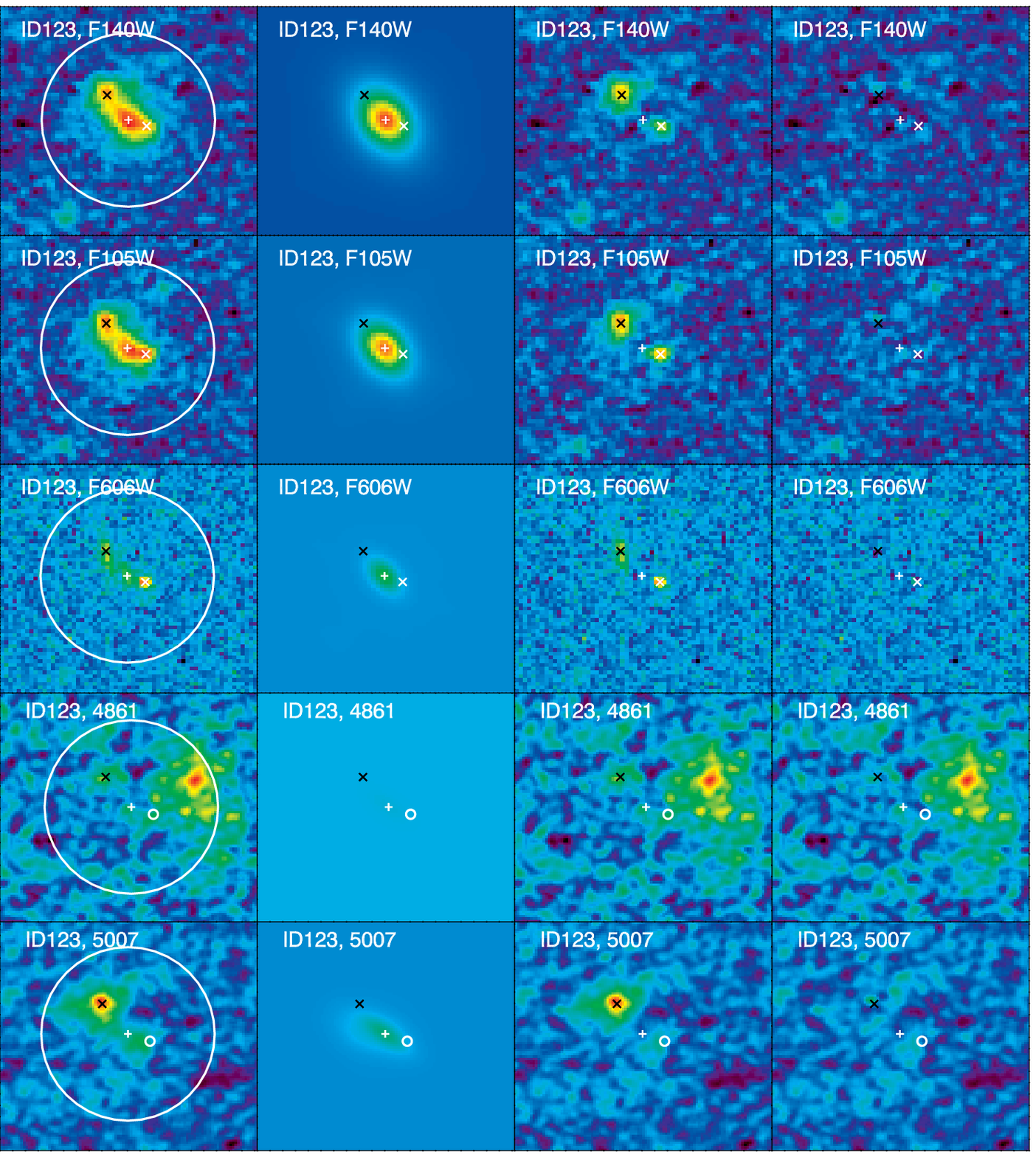}
\vspace{-3cm}
  \caption{Continuum and emission line maps the galaxy ID123. From \textit{top to bottom}: F140W, F105W, F606W continuum maps, [OII], H$\beta$, [OIII], and/or H$\alpha$ emission line maps (the filter or wavelength of the map is indicated in the top left corner of each panel). From \textit{left to right:} continuum or emission line map, GALFITM model for the diffuse component (single S\'ersic profile), residuals obtained subtracting the model (column 2) from the original image (column 1), residuals obtained subtracting GALFITM best-fit model (including the diffuse S\'ersic profile plus additional PSFs at the location of the blobs detected in the residuals shown in column 3) from the original map (column 1). The crosses indicate the center of the diffuse S\'ersic profile. The ``x'' and small empty circles indicate the center of the blobs (respectively detected with S/N $>$ 3 or non-detected); white and black are used to indicate blobs with PSF and S\'ersic profile. The large white circle indicates the area considered when finding the blobs (and corresponding to 5 galaxy effective radii). We notice that the [OIII]4959$\mathrm{AA}$ is not subtracted in these maps, so it appears as a ``ghost'' emission in the maps labelled as ``5007'' (but in the analysis we correct the [OIII]5007$\mathrm{AA}$ fluxes for its contribution, Section \ref{subsec:emission_line_maps}). Similarly the [OIII]5007$\mathrm{\AA}$ is not subtracted from the H$\beta$ maps and therefore it appears as bright blobs in the righten side of most of the maps labelled as ``4861''. In the following we report the same figure for each galaxy in our sample (the ID is reported in the top left corner of each panel). Each stamp has a size of 3.7''$\times$3.7'' ($\sim 30 \times 30$ kpc at $z \sim 1$ -- 3), we adopt the same color bar in all the panels and an inverse hyperbolic sine scaling. \textit{The complete version of this figure is available online.}}
   \label{fig:panels_stat_sample1}
\end{figure*}

We report below the results that we obtained fitting the light profiles of our sample galaxies with the algorithm GALFITM, according to the procedure described in Section \ref{subsec:blobs_flux}. We show the continuum and emission line maps of each galaxy, together with the 2D best-fit model of their light profile and the residual maps in Figures from \ref{fig:panels_stat_sample1} to \ref{fig:panels_stat_sample_last}. We report the physical properties of our sample galaxies in Table \ref{tab:properties_gal}. Finally, we report the physical properties of the compact and extended blobs in Table \ref{tab:properties_blobs}.


\clearpage

\section{Estimate of the flux uncertainties and sample completeness}
\label{app:uncertainties}

To estimate the uncertainties associated to the flux estimate of the blobs, we ran 1000 Monte Carlo simulations. We injected one fake PSF at the time in the continuum and emission line maps of our galaxies, within 3 effective radii, avoiding any overlap with already existing blobs. The main issue we wanted to understand with these simulations was how well GALFITM retrieved PSFs on top of a disk. Therefore, not to introduce further degrees of freedom and degeneracies in the modelling, we considered the best-fit GALFITM model obtained for each galaxy before the injection of the fake PSF, we kept its structural parameters fixed (allowing only the magnitude of each component to change), and we added to this baseline model a PSF profile. The initial guesses for the magnitude and center coordinates of the additional PSF were estimated by randomly perturbing the known, input values of the simulation. We did not visually inspect all the residuals after the subtraction of the best-fit model, but we defined the following automatic criteria, calibrated with the real data, to decide when a fit failed. With SExtractor we created a segmentation map for each galaxy and we looked, within this region, at the variations of the residuals normalized by the input image. This was done so that neighbouring noisy pixels would not be included in the fit. We computed the mean ($M$) and standard deviation ($S$) of the distribution of the normalized residuals after 3$\sigma$ clipping and considered a fit as reliable if the following conditions were simultaneously satisfied: $S_{\mathrm{continuum}} \leq 0.3$, $S_{\mathrm{maps}} \leq 0.4$, $|M_{\mathrm{continuum}}| \leq 3S_{\mathrm{continuum}}$, and $|M_{\mathrm{maps}}| \leq 3S_{\mathrm{maps}}$ (i.e. variations between pixels $<$ 30\% -- 40\%, non-structured positive/negative residuals). Visual inspection of the residuals of random simulated galaxies confirmed that our automatic criteria were properly selecting the good fits (e.g. the ones with small and smooth residuals), allowing us to exclude the non reliable ones ($\sim$ 5\% of all the fits) in order not to pollute our uncertainties estimate.

To determine the uncertainties associated with the flux of the blobs, we divided the simulated PSFs in bins based on their distance from the galaxy barycenter and the magnitude of the underlying disk as measured by GALFITM on the real data. For each bin, we computed the difference between the known input flux of the fake injected PSF and the one retrieved by GALFITM. The standard deviation of the sigma clipped distribution of these differences gave us, in each magnitude and distance bin, the flux uncertainty. It was important to divide blobs in bins of galaxy magnitude and distance from the nucleus, since the ability of GALFITM to correctly retrieve the flux is highly dependent on the contrast of the PSF with respect to the underlying diffuse disk. If the disk is bright (faint) the contrast is low (high) and therefore the uncertainties are larger (smaller). Furthermore, if the blob is close to the nucleus of the galaxy the contrast is low due to the rise of the S\'ersic profile of the disk and the uncertainties are large. On the contrary, if the blob is located in the outskirts of the galaxy the contrast is high and the uncertainties are quite small. This is indeed the trend that we find with our simulations. We finally interpolated all the bins to get the uncertainties as a function of the galaxy magnitude and the blob distance from the center. We imposed that at large distances from the nucleus, where the contrast is very high and the uncertainty derived with our simulations is negligible, the flux error was set by the background limiting magnitude, estimated with aperture photometry on empty regions of the sky. In the emission line maps the underlying galaxy disk was usually very faint, thus the uncertainties associated to the emission line flux of the blobs are simply set by the background limiting magnitude, without any dependence on their distance from the galaxy barycenter or disk brightness.

In our sample of blobs we also included the extremely young clump dubbed ``Vyc1'' (\citealt{Zanella2015}) and we analyzed it with the automatic procedure presented in Section \ref{subsec:finding_blobs} and \ref{subsec:blobs_flux}. To further check the correctness of our automatic procedure in retrieving the fluxes of the blobs, we compared the emission lines and continuum flux of Vyc1 with the values that we obtained with the careful and customized analysis presented in \cite{Zanella2015}. We found completely consistent results.

Finally, we used our Monte Carlo simulations to estimate the uncertainties associated to the coordinates of the blobs, using an analogous method to the one presented for the flux uncertainties.

\begin{figure*}
	\includegraphics[width=0.8\textwidth]{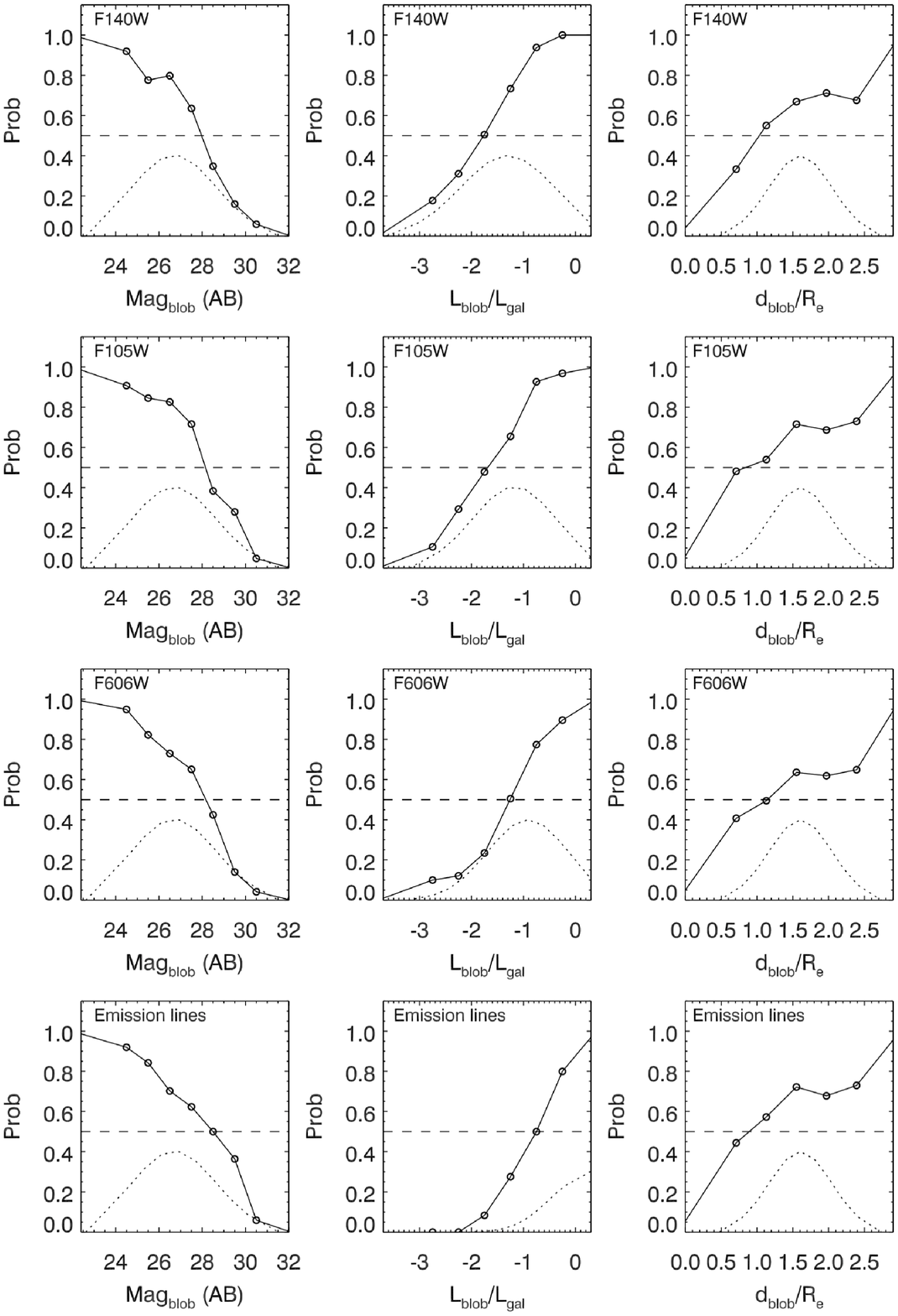}
  \caption{Detection probability of blobs. \textit{Columns (from left to right):} probability to detect fake blobs as a function of their magnitude, the ratio between the blob and galaxy luminosity, and their distance from the galaxy barycenter normalized by the galaxy effective radius (empty symbols connected by solid curves). \textit{Rows (from top to bottom):} continuum F140W filter, F105W filter, F606W filter, and emission line maps. In each panel we also report the detection probability of 50\% (dashed line) and the distribution of the input parameters of the fake blobs (dotted curves). In particular, the distribution of the initial parameters of fake blobs injected in emission line maps results shifted towards high values of L$_{\mathrm{blob}}/$L$_{\mathrm{galaxy}}$. This is due to the fact that the galaxies are typically fainter in the emission line rather than in the continuum maps.}
   \label{fig:completeness_functions}
\end{figure*}

To estimate completeness of the sample we used the same Monte Carlo simulations described above. We divided the sample of mock blobs that we injected on top of real galaxies in bins based on their intrinsic flux ($F_{\mathrm{input}}$). For each bin we estimated the fraction of blobs whose flux was retrieved by GALFITM ($F_{\mathrm{output}}$) with a relative uncertainty smaller than 50\%: $(F_{\mathrm{input}} - F_{\mathrm{output}})/F_{\mathrm{input}} \leq 0.5$. As expected, the fraction of blobs that satisfy this condition decreases as their magnitude increases. Our sample is 50\% complete down to 28 mag in F140W, 28.3 mag in F105W, 28.1 mag in F606W imaging, and 28.8 mag in emission line maps (Figure \ref{fig:completeness_functions}).

\section{Distribution of observed blobs properties}
\label{app:observed_properties}

\begin{figure*}
	\includegraphics[width=\textwidth]{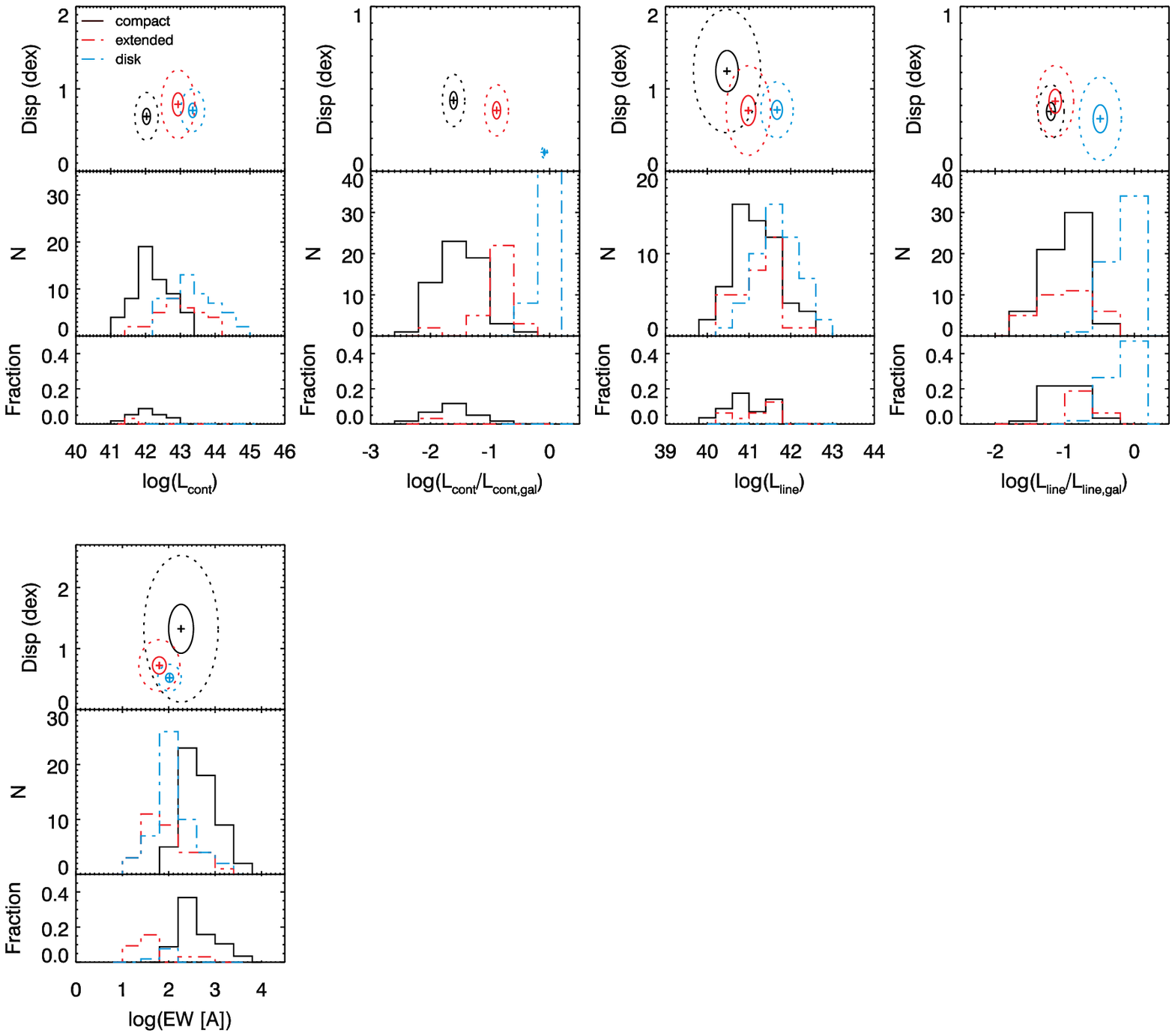}
  \caption{Distribution of the observed properties of blobs and disks. Compact blobs are shown in black, extended ones in red, and the underlying disks in cyan. \textit{Top panels (from left to right):} we show the distribution of continuum luminosity, continuum luminosity normalized by the total galaxy continuum luminosity, emission line luminosity, and emission line luminosity normalized by the total galaxy line luminosity. \textit{Bottom panels (from left to right):} we show the distribution of equivalent width. For each observable we show in the \textit{middle panel} the distribution of the parameters, in the \textit{bottom panel} the fraction of upper or lower limits in each bin, and in the \textit{top panel} the mode and dispersion (cross) of each distribution (solid and dotted ellipses indicate the 1$\sigma$ and 3$\sigma$ uncertainty on the mode and dispersion), computed as described in Section \ref{subsec:distributions}.}
   \label{fig:distr_obs}
\end{figure*}

\begin{figure*}
	\includegraphics[width=\textwidth]{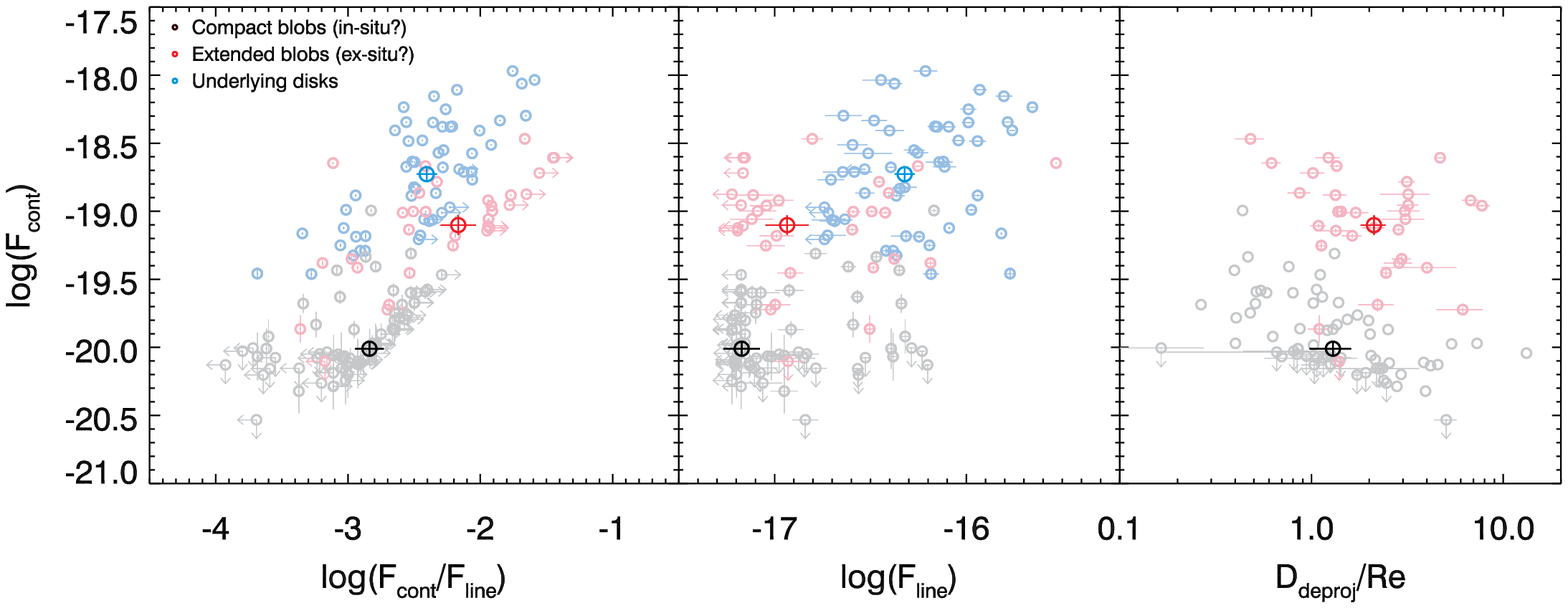}
  \caption{Observable properties of blobs and disks. In each panel, we show the relation between continuum flux and continuum-to-emission line flux ratio (\textit{left panel}), emission line flux (\textit{central panel}), deprojected distance from the galaxy barycenter (\textit{right panel}). In the background we show individual compact (black symbols) and extend (red  symbols) blobs, and the underlying disks (cyan symbols), whereas the foreground symbols show the mean properties of each population.}
   \label{fig:observables_trends}
\end{figure*}

\clearpage
\onecolumn
\tiny
\begin{landscape}
\begin{longtable}{cccccccccccccccc} 
	\caption{Properties of our sample galaxies. \textit{The full table is avialable online.}}
	\label{tab:properties_gal}
\\ \toprule
\midrule
ID  &  R.A. &    Decl. &   Redshift &    Re &   $\log($M$_\star)$ &   $\log($SFR$)$ &  E(B-V)  &    Metall. &  F140W  &  F105W &   F606W  & [OII] &   Hb  &     [OIII]    &  Ha      \\     
   &  [Deg]  &   [Deg]   &   &   [kpc]   &       [$\log(\mathrm{M_\odot})$]   &   [$\log(\mathrm{M_\odot yr^{-1}})$]         &   &   &   [$10^{-21}$cgs]   & [$10^{-21}$cgs] & [$10^{-21}$cgs] & [$10^{-17}$cgs] & [$10^{-17}$cgs] & [$10^{-17}$cgs] & [$10^{-17}$cgs] \\
(1) & (2) & (3) & (4)  & (5) & (6) & (7) & (8) & (9) & (10) & (11) & (12) & (13) & (14) & (15) & (16) \\
\midrule
ID107 &  222.31832273 &  8.92183343 &  2.534$\pm0.002$ &  1.7 &  10.2 &   2.1  & 0.3$\pm0.2$ & --  &    2.69$\pm  0.02$ &    2.69$\pm  0.02$ &    4.60$\pm  0.05$  &   5.56$\pm  0.74$ &  --& -- &  --\\
ID123 &  222.32099014 &  8.92272076 &  1.877$\pm0.003$ &  2.7 &   9.6 &   1.5  & 0.1$\pm0.2$ & --  &    1.71$\pm  0.02$ &    1.73$\pm  0.03$ &    3.21$\pm  0.09$  & --&    $<$0.36 &   1.97$\pm   0.34$ &  --\\
ID151 &  222.30327383 &  8.92363322 &  1.721$\pm0.002$ &  1.8 &   9.9 &   1.8  & 0.3$\pm0.2$ & --  &    2.10$\pm  0.02$ &    2.46$\pm  0.02$ &    3.46$\pm  0.04$  & --&    1.36$\pm  0.19$ &   4.04$\pm   0.18$ &  --\\
\bottomrule
\end{longtable}
\begin{minipage}{23.5cm}
Column (1) Galaxy ID; (2) Right ascension; (3) Declination; (4) Redshift; (5) Effective radius. Typical uncertainties are 20\%; (6) Stellar mass estimated from the SED fitting (Section \ref{subsec:galaxy_properties}). Typical uncertainties are $\sim$ 0.2 dex; (7) Star formation rate from SED fitting. Typical uncertainties are $\sim$ 0.2 dex; (8) Reddening estimated from the Balmer decrement when longslit Subaru/MOIRCS spectre are available, from SED fitting otherwise (Section \ref{subsec:galaxy_properties}); (9) Gas phase metallicity 12 + log(O/H) estimated from the \oiii/\oii line ratio, when the lines were detected (Section \ref{subsec:metallicity}); (10) Observed galaxy flux in the F140W broadband filter; (11) Observed galaxy flux in the F105W broadband filter; (12) Observed galaxy flux in the F606W broadband filter; (13) Observed \oii ~flux of the galaxy; (14) Observed H$\beta$ flux of the galaxy; (15) Observed \oiii ~flux of the galaxy; (16) Observed H$\alpha$ flux of the galaxy. 
\end{minipage}
\end{landscape}

\clearpage
\onecolumn
\footnotesize
\begin{landscape}
\begin{longtable}{ccccccccccccc}  
  \caption{Physical properties of the compact and extended blobs. \textit{The full table is available online.}}
  \label{tab:properties_blobs}
\\ \toprule
\midrule
ID$_\mathrm{host}$ &  c/e &  F140W   &  F105W  &  F606W           &      [OII]  &  Hb &   [OIII] &   Ha &     M$_\star$ &  SFR &  Age &  Metall.          \\
            &                     & [$10^{-21}$cgs]   & [$10^{-21}$cgs]   & [$10^{-21}$cgs]   & [$10^{-17}$cgs]   & [$10^{-17}$cgs]   & [$10^{-17}$cgs]   & [$10^{-17}$cgs]   & [$10^{8}\mathrm{M_\odot}$]   &   [$\mathrm{M_\odot yr^{-1}}$] &  [$10^{7}$yr] & \\
(1) & (2) & (3) & (4)  & (5) & (6) & (7) & (8) & (9) & (10) & (11) & (12) & (13) \\
\midrule
ID214 &  c &    9.76$\pm  2.77$ &    $<$4.24 &  $<$19.76 &    $<$0.22 &  -- &  -- &  -- &     8.29$\pm  3.86$  &   $<$20.2  &   $>$1.14 &  -- \\
ID279 &  c &   12.02$\pm  3.00$ &   47.67$\pm  4.08$ &  170.18$\pm 13.95$ &    $<$0.42 &    $<$0.21 &    4.79$\pm  0.20$ &  -- &    --  &   --  &   -- &  -- \\
ID279 &  c &    $<$2.77 &   14.40$\pm  4.14$ &  $<$19.03 &    $<$0.42 &    $<$0.21 &    0.80$\pm  0.20$ &  -- &     --  &   --  &    -- &  -- \\
ID279 &  c &    $<$2.17 &    $<$3.69 &  $<$14.04 &    $<$0.42 &    $<$0.21 &    0.84$\pm  0.20$ &  -- &     --  &   --  &    -- &  -- \\
\bottomrule
\end{longtable}
\begin{minipage}{23.5cm}
Column (1) ID of the galaxy hosting the blob; (2) Compact ``c'' or extended ``e'' blob; (3) Observed blob flux in the F140W broadband filter; (4) Observed blob flux in the F105W broadband filter; (5) Observed blob flux in the F606W broadband filter; (6) Observed \oii ~flux of the blob; (7) Observed H$\beta$ flux of the blob; (8) Observed \oiii ~flux of the blob; (9) Observed H$\alpha$ flux of the blob; (10) Stellar mass (Section \ref{subsec:stellar_mass}); (11) Star formation rate (Section \ref{subsec:sfr}); (12) Age (Section \ref{subsec:blobs_age}); (13) Gas phase metallicity 12 + log(O/H) estimated from the \oiii/\oii line ratio, when the lines were detected (Section \ref{subsec:metallicity}).
\end{minipage}
\end{landscape}

\twocolumn
\section{Spatially-resolved stellar mass maps}
\label{app:mass_maps}

\begin{figure*}
	\includegraphics[width=\textwidth]{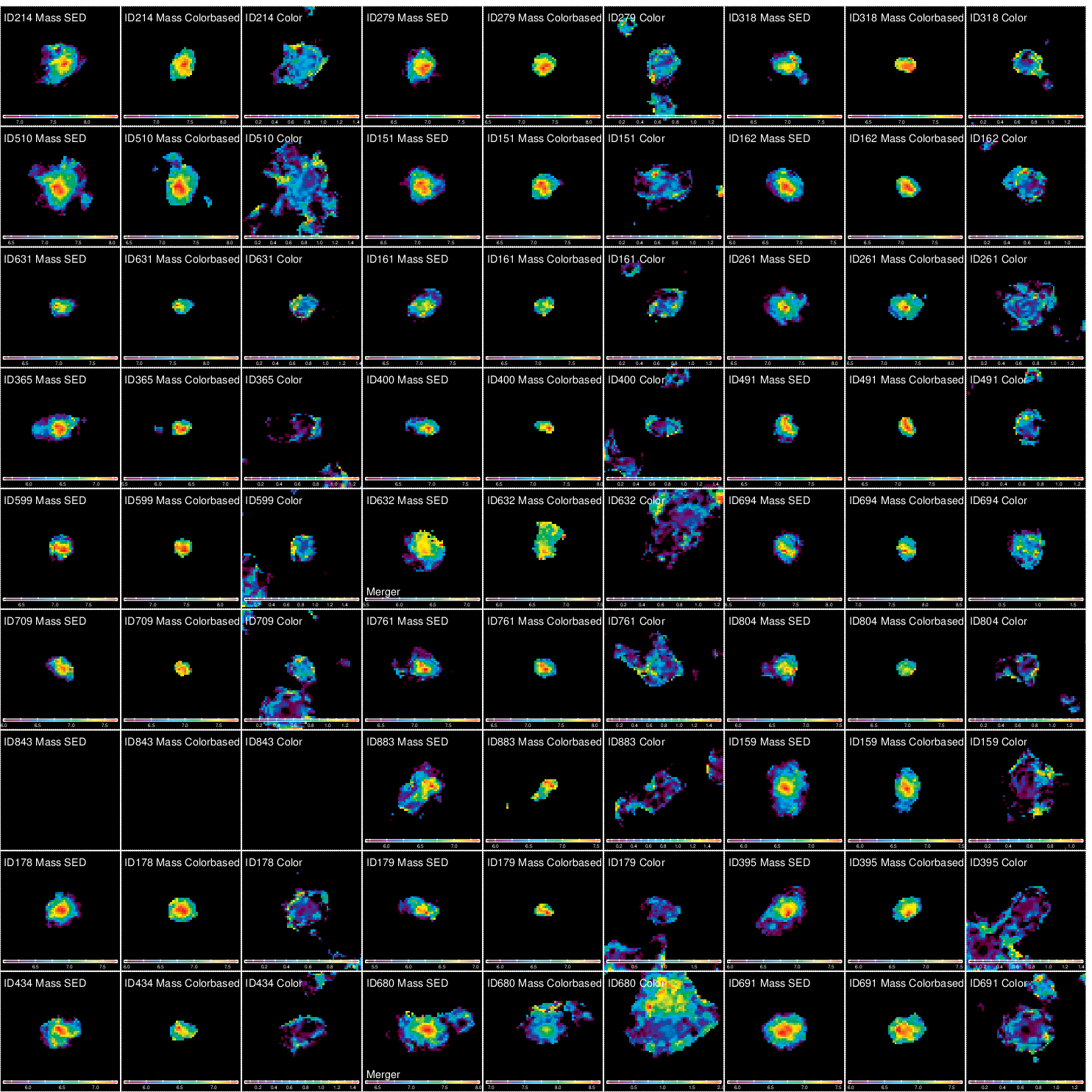}
  \caption{Spatially-resolved mass and color maps of our sample galaxies. From \textit{left to right}, in each column: stellar mass map estimated from pixel-to-pixel SED fitting (in units of $\log \mathrm{M_\star}$); color-based stellar map map (in units of $\log \mathrm{M_\star}$); color (F140W, F105W) color (in units of AB mag). Each stamp has a size of 3.7''$\times$3.7''. \textit{A complete figure is available online.}}
   \label{fig:mass_maps}
\end{figure*}

By using the three available \textit{HST} continuum bands (F140W, F105W, F606W) we have reconstructed the stellar mass distribution of galaxies in two ways: on one hand we have performed spatially-resolved pixel-to-pixel SED fitting, on the other we have estimated stellar masses based on galaxies' color (based on the F140W and F105W magnitudes, \citealt{Wuyts2013}, \citealt{Cibinel2015}). The two estimates are in good agreement (Figure \ref{fig:mass_maps}). To classify the galaxies into major mergers and non-interacting we considered the non-parametric structural parameters asymmetry (namely the normalized residual flux as obtained from the difference between the original image and its 180$^{\circ}$-rotated version, \citealt{Conselice2003}) and M$_\mathrm{20}$ (namely the normalized second-order moment of the 20\% brightest pixels, \citealt{Lotz2004}) measured on the resolved stellar mass maps as done by \cite{Cibinel2015} and \cite{Cibinel2018}. Measuring these parameters on the stellar mass maps rather than using single-band images reduces the misclassification of clumpy, non-interacting galaxies. The clumps in fact are bright in rest-frame UV imaging, but do not contribute substantially to the total mass budget of the host galaxy. The stellar mass maps of clumpy disks appear smooth and symmetric, at odds with their single-band UV images. Major mergers instead enclose a significant fraction of the mass of the host galaxy ($\gtrsim$ 1:4 -- 1:5) and therefore produce variations in the asymmetry and M$_{20}$ parameters \citep{Cibinel2015}. The mass-based selection of mergers has been calibrated on the MIRAGE hydrodynamical numerical simulations of isolated and merging galaxies \citep{Perret2014}.

\bsp	
\label{lastpage}
\end{document}